\documentclass[%
 reprint,
superscriptaddress,
 amsmath,amssymb,
 aps,
]{revtex4-2}

\usepackage{graphicx}
\usepackage{dcolumn}
\usepackage{bm}
\usepackage{xspace}
\usepackage{color}
\usepackage{ulem}
\usepackage[unicode,breaklinks,bookmarks=true]{hyperref}

\usepackage{soul}
\usepackage[dvipsnames]{xcolor}

\def\lesssim{\mathrel{\hbox{\rlap{\hbox{\lower4pt\hbox{$\sim$}}}\hbox{$<$}}}}
\def\gtrsim{\mathrel{\hbox{\rlap{\hbox{\lower4pt\hbox{$\sim$}}}\hbox{$>$}}}}

\newcommand{\bea}{\begin{eqnarray}}
\newcommand{\eea}{\end{eqnarray}}

\newcommand{\bP}{{\bf P}}
\newcommand{\bF}{{\bf F}}
\newcommand{\bU}{{\bf U}}

\newcommand{\prim}{{{\mathbf{P}}}}

\newcommand{\harm}{\xspace{\sc Harm3d}\xspace}

\def\lambdabar{%
\relax
\bgroup
\def\@tempa{\hbox{\raise.73\ht0
\hbox to0pt{\kern.25\wd0\vrule width.5\wd0
height.1pt depth.1pt\hss}\box0}}%
\mathchoice{\setbox0\hbox{$\displaystyle\lambda$}\@tempa}%
{\setbox0\hbox{$\textstyle\lambda$}\@tempa}%
{\setbox0\hbox{$\scriptstyle\lambda$}\@tempa}%
{\setbox0\hbox{$\scriptscriptstyle\lambda$}\@tempa}%
\egroup
}

\begin{document}

\preprint{APS/123-QED}

\title{Superposed metric for spinning black hole binaries approaching merger}

\author{Luciano Combi}
\email{lcombi@iar.unlp.edu.ar} 
\affiliation{Instituto Argentino de Radioastronom\'ia (IAR, CCT La Plata, CONICET/CIC), C.C.5, (1984) Villa Elisa, Buenos Aires, Argentina}
\affiliation{%
 Center for Computational Relativity and Gravitation,
  Rochester Institute of Technology, Rochester, NY 14623.\\
}%
\author{Federico G. Lopez Armengol}%

\author{Manuela Campanelli}
\affiliation{%
 Center for Computational Relativity and Gravitation,
  Rochester Institute of Technology, Rochester, NY 14623.\\
}%

\author{Brennan Ireland}
\affiliation{%
 Center for Computational Relativity and Gravitation,
  Rochester Institute of Technology, Rochester, NY 14623.\\
}%
\affiliation{%
The US Agency for International Development - 1300 Pennsylvania ave NW, Washington, DC, 20004\\
}%

\author{Scott C. Noble}
\affiliation{
 Gravitational Astrophysics Laboratory, Goddard Space Flight Center, Greenbelt, MD 20771.\\
}%

\author{Hiroyuki Nakano}
\affiliation{%
 Faculty of Law, Ryukoku University, Kyoto 612-8577, Japan\\
}%

\author{Dennis Bowen}
\affiliation{
Center for Theoretical Astrophysics, Los Alamos National Laboratory,
  P.O. Box 1663, Los Alamos, NM 87545.\\
}%
\affiliation{
 X Computational Physics, Los Alamos National Laboratory,
  P.O. Box 1663, Los Alamos, NM 87545.
}%

\date{\today}

\begin{abstract}
We construct an approximate metric that represents the spacetime of spinning binary black holes (BBH) approaching merger. We build the metric as an analytical superposition of two Kerr metrics in harmonic coordinates, where we transform each black hole term with time-dependent boosts describing an inspiral trajectory. The velocities and trajectories of the boost are obtained by solving the post-Newtonian (PN) equations of motion at 3.5 PN order.  We analyze the spacetime scalars of the new metric and we show that it is an accurate approximation of Einstein's field equations in vacuum for a BBH system in the inspiral regime. Furthermore, to prove the effectiveness of our approach, we test the metric in the context of a 3D general relativistic magneto-hydrodynamical (GRMHD) simulation of accreting mini-disks around the black holes. We compare our results with a previous well-tested spacetime construction based on the asymptotic matching method. We conclude that our new spacetime is well-suited for long-term GRMHD simulations of spinning binary black holes on their way to the merger.

\begin{description}
\item[Usage]
Secondary publications and information retrieval purposes.
\item[Structure]
You may use the \texttt{description} environment to structure your abstract;
use the optional argument of the \verb+\item+ command to give the category of each item. 
\end{description}
\end{abstract}

\maketitle


\section{\label{sec:intro} Introduction}

There is abundant evidence that most galaxies harbor supermassive black holes (SMBHs) at their centers \cite{Gultekin09, volonteri2010formation}. A non-negligible fraction of these galaxies undergoes one or more mergers within a Hubble time \cite{Hopkins_2010, Naab_2017}. After two galaxies merge, various processes, such as dynamical friction, might drive their SMBHs to close separations of sub-parsec scales \citep{Khan16, Kelley17, Pfister:2017uxn}. 

In this situation, the binary system starts emitting gravitational radiation efficiently, losing energy, and eventually merging  \citep{PhysRevLett.95.121101,
PhysRevLett.96.111101, PhysRevLett.96.111102}. The frequencies of these gravitational waves (GWs) span from nano-Hertz for the inspiral phase \cite{burke2019astrophysics}, up to milli-Hertz for the merger. Pulsar timing array consortiums and LISA are actively working towards detecting these GW for the first time in the next decade \citep{Del_Pozzo_2018, perera2019international, baker2019laser, mcwilliams2019astro2020}.

Because galaxy mergers can be very efficient at driving interstellar gas toward the galactic center \cite{Chapon_2013, Tremmel_2017,Hopkins_2010}, SMBH binary mergers should accrete enough gas and emit observable electromagnetic (EM) radiation \citep{Colpi14}.  The total energy radiated is proportional to the gas mass present during the SMBH binary merger \cite{Krolik_2010}, and its characteristic form of this EM emission is expected to be distinct relative to ordinary accreting supermassive black hole holes, such as a single active galactic nucleus (AGN) \citep{Roedig:2014,2015Natur.525..351D, Krolik2019}. Identifying those signatures, however, requires complex calculations of the SMBH binary mergers and their associated luminosity, spectrum, and time-dependence.  Some numerical calculations have started to reveal interesting properties associated with these EM signals \citep{Bowen17, RyanMacFadyen17, Bowen18,dAscoli:2018fjw,2018MNRAS.476.2249T}. However, because they depend strongly on the system's properties such as the total binary mass, mass-ratio, spins (magnitude and direction), and accretion rate, much work remains to be done.

Since the interstellar gas of the merged galaxies is expected to have a considerable amount of angular momentum, a circumbinary disk will form around the BHs \cite{Springel2005}. Semi-analytical models of these systems predicted that the binary would decouple from the fluid and coalesce in a \textit{dry} merger; in other words, the inflow time of the accreted matter would become larger than the inspiral time at short orbital separations \citep{MP05,Pringle1991}. 

Since the equations of magneto-hydrodynamics (MHD) in dynamical spacetimes are highly non-linear, we need numerical simulations to make accurate predictions. In the past decade, $\alpha$-viscous simulations and 3D MHD simulations demonstrated that accretion onto the binary occurs even in the late inspiral phase \citep{MM08,2010ApJ...715.1117B,Pal10,Farris11,Bode12,Farris12,Giacomazzo12,Noble12,Shi12,DOrazio13,Farris14,Gold14,Farris15,Farris15a,Shi2015,DOrazio16,Bowen18,2018MNRAS.476.2249T}, with appreciable accretion sustained right up to the time of merger when using relativistic inspiral rates \citep{Noble12}. These simulations also showed that the circumbinary disk is truncated at a distance approximately twice the binary separation from the system's center-of-mass. Outside this truncation radius, mass piles up, forming a local peak in the surface density profile; inside this radius, the accretion flow onto the binary is confined within two narrow streams traversing a low-density gap. Each of these streams is associated with one of the BHs, forming mini-disks around each hole.

Moreover, most of these simulations revealed the formation of a characteristic $m=1$ mode overdensity, or \textit{lump}, in the circumbinary disk for mass-ratios close to unity \citep{MM08,Noble12,Shi12,DOrazio13,Farris14,Farris15,Farris15a,DOrazio16,2018MNRAS.476.2249T}. In that case, the lump modulates the accretion of the system and feeds the BHs with a single-arm stream. If the mass ratio is small, the lump is weak \cite{noble2020}, and the lighter BH receives most of the mass, carving a path near the inner edge of the circumbinary disk. 

Solving Einstein's field equations for the metric of SMBH binaries coupled to MHD fields is computationally challenging. This has been done in the past in the force-free regime \cite{alic2012accurate, moesta2012detectability, palenzuela2010dual, palenzuela2010magnetospheres}, or for close binary separations \cite{Giacomazzo12,Farris12, Farris14a, Gold_2014a, Gold_2014b}. Recently, authors in Ref. \cite{paschalidis2021minidisk} have evolved a spinning BBH in full GRMHD for a few orbits in the inspiral regime, focusing on mini-disks dynamics.

In order to model realistic scenarios in these systems, however, we need first to evolve the circumbinary disk for many binary orbits until reaching a steady-state \cite{armengol2021circumbinary}. In particular, the presence of the $m=1$ lump mode in the circumbinary disk is very important to determine the dynamics of what happens in the inner cavity \citep{bowen2019}. An alternative approach to achieve this is to use approximate, semi-analytical, solutions of Einstein's equations for spacetime, and evolve the MHD equations on it. This allows one to explore the parameter space of the spacetime more efficiently and to focus computational resources (such as the configuration of the grid) on the MHD fields. As an example of this approach, previous work evolved relativistic circumbinary disks with a post-Newtonian (PN) metric during the inspiral regime \citep{noble2012, zilhao2014pn, noble2020}. Since this metric is only valid far from the BHs, these simulations must excise the binary region from the computational domain.

To analyze the strong-field behavior of the plasma near the BHs, we need a background metric that is valid at these scales. Such metric can be built, for instance, through the so-called asymptotic matching approach \citep{Mundim:2013vca,Nakano:2016klh,ireland2016} that stitches different known analytical approximations for a binary black hole (BBH) metric. In this approach, a perturbed Schwarzschild or Kerr solution is used for the inner-zone, a PN expansion for the near-zone, and a post-Minkowskian expansion for the far-zone are glued together via the transition techniques developed in Refs. \cite{yunes2005,yunes2006}.  This metric has been used to perform GRMHD simulations of accretion flows with mini-disks around non-spinning BHs for the first time \cite{bowen2017,bowen2018,bowen2019}. 

This approach for the spacetime construction can be generalized to spinning BHs \citep{Gallouin:2012kb,Nakano:2016klh,ireland16}, but the analytical matching metric becomes too complex and computationally expensive for long-term GRMHD simulations. This motivates the search for more efficient approaches for building an analytical spinning BBH metric. From numerical relativity simulations, we know that spins play a key role in BBH inspirals and mergers. Spins aligned with the orbit, for example, can significantly alter the pace of orbital evolution by gravitational radiation~\cite{Campanelli:2006uy,  hemberger2013final,healy2018hangup}. Oblique spins can drive complex precession and nutation whose amplitude increases rapidly at smaller separations \cite{Campanelli:2006fy}; spins with partial orbital alignment can repeatedly flip sign \cite{Lousto:2014ida, Lousto:2015uwa, lousto2016spin, Lousto:2018dgd}; spin-orbit PN resonances can also tilt the orbital orientation \cite{kesden2014gravitational}.

In Ref. \cite{armengol2021circumbinary}, we introduced a new approach for spinning BBHs, building the approximate metric as a linear superposition of two boosted Kerr-Schild BHs. We used this new approximate spacetime to analyze the accretion of a circumbinary disk around the BBH in a Keplerian orbit and unveil the influence of the spin in the circumbinary flow. We found that streams falling into the binary cavity as well as the accretion rate are affected by the magnitude and direction of the BH spins, while other properties in the bulk of the circumbinary disk remain unaffected. In particular, due to frame dragging effects, accretion decreases (increases) when the spins are (anti-)aligned, with important effects on the overall luminosity of the system. We also find that the circumbinary disk is stabilized after more than 100 orbits, which implies that long simulations are required for making realistic predictions. In the present work, we formalize and extend this previous approach by superposing two boosted Kerr BHs in harmonic coordinates, solving the PN equations of motion (EOM) for the BH trajectories. This allows us to have a more accurate approximation for the spacetime and to analyze the influence of the inspiral on matter orbiting the BBHs. We test this metric by analyzing its spacetime scalars and using it in a full 3D GRMHD simulation of accreting mini-disks. We compare our results with the more expensive and complex matching metric for non-spinning BHs. Having tested the viability and accuracy of our spacetime, in an upcoming work we will use our new metric for analyzing the influence of spins in the mini-disk dynamics and their outflows. 

We organize the paper as follows: in Section II, we build the approximate metric in the harmonic gauge by boosting and superposing two BH terms. In Section III, we test the metric by analyzing its spacetime scalars and comparing them with the asymptotic matching approach. In Section IV, we test the metric as a background spacetime for a GRMHD simulation, comparing again with previous results from the asymptotic matching approach. We conclude that the metric is accurate and robust to be used in accretion disk simulations of BBH in the inspiral regime.

\textbf{Notation and conventions.} We use the signature $(-,+,+,+)$ and we follow the Misner-Thorne-Wheeler convention for tensor signs. We use geometrized units, $G=c=1$. We use Latin letters $a,b,c,...=0,1,2,3$ for four dimensional components of tensors, and $i,j,k,...=1,2,3$ for space components. An orthonormal space basis is written as $\vec{e}_{(i)}= e^{a}_{(i)} \partial_{a}$, where its components are denoted as $(i),(j),(k),...=1,2,3$.

\section{Construction of superposed binary black holes metric}

To solve Einstein's field equations with numerical methods, one usually starts from a three-dimensional slice of the spacetime metric and matter fields as initial data for the problem \cite{baumgarte, YorkCTT1971, BowenYorkSolutions1980}. This initial data cannot be arbitrary since it has to satisfy the constraints of Einstein's equations; there is, however, a significant amount of freedom to choose it because the equations are invariant under diffeomorphisms. In General Relativistic simulations, some of these choices are preferred over others by their numerical robustness and accuracy. For instance, in the Extended Conformal Thin Sandwich formalism \cite{york1999, pfeiffer2003} used by the Spectral Einstein Code (SpEC) \cite{spec}, we can freely specify the conformal metric, the trace of the extrinsic curvature, and their time derivatives. For BBH simulations, a widely used and well-tested choice for the conformal metric is a superposition of two BHs in Kerr-Schild coordinates \cite{matzner, lovelace2008, cook2004, varma2018}.

Motivated by this approach, we shall test an ansatz for a 4-dimensional BBH metric constructed as a superposition of the form $g_{ab}(t) \sim \eta_{ab} + H^{(1)}_{ab}(t)+H^{(2)}_{ab}(t)$, where $\eta_{ab}$ is the Minkowski background metric and the terms $H^{(n)}_{ab}(t)$, $n=1,2$, correspond to each BH. Each BH term is boosted with a time-dependent transformation following the trajectories of the holes. These trajectories can be accurately described by solving the PN equations of motion \cite{Blanchet:2013haa} for orbital separations larger than $\sim 10\: M$. The final BBH metric is a simple, time-dependent, analytical function that we can use as a background spacetime for MHD simulations. In this section, we show how to build this ansatz.

\subsection{Kerr black hole in harmonic coordinates}

The metric of a spinning BH in Kerr-Schild (KS) coordinates, $\lbrace t_{\rm KS},\,x_{\rm KS},\,y_{\rm KS},\,z_{\rm KS} \rbrace$,  is the natural choice for building a superposed metric since it has the form of a background term plus a BH term (see Ref. \cite{poisson,visser2007kerr}):
\begin{equation}
    g_{ab} = \eta_{ab} + 2 H l_{a} l_{b},
    \label{eq: ks}
\end{equation}
where $\eta_{ab}$ is the Minkowski metric in  Cartesian coordinates, and the null covector $l_a$ is defined as
\begin{align}
    -l_{a}dx^a_{\rm KS} := & dt_{\rm KS}+ \frac{r \: x_{\rm KS} + a \: y_{\rm KS}}{r^2+a^2}dx_{\rm KS}
\cr &
+ \frac{r \: y_{\rm KS}-a \: x_{\rm KS}}{r^2+a^2}dy_{\rm KS} + \frac{z_{\rm KS}}{r}dz_{\rm KS},
\end{align}
with the spin parameter $a$, the function:
\begin{equation}
    H:= \frac{2Mr^3}{r^4+a^2 z_{\rm KS}^2},
\end{equation}
where $M$ is the mass of the black hole, and the Boyer-Lindquist radius $r$ is given by
\begin{equation}
    r^2:= \frac{1}{2}( r_{\rm KS}^2-a^2) \left( 1 + \sqrt{1+\frac{4 a^2 z_{\rm KS}^2}{(r_{\rm KS}^2-a^2)^2}} \right),
    \label{eq: blradius}
\end{equation}
where
\begin{equation}
r_{\rm KS}^2:= x_{\rm KS}^2+y_{\rm KS}^2+z_{\rm KS}^2.    
\end{equation}

This coordinate system was used in our previous work \cite{armengol2021circumbinary} to build an approximate spacetime metric for a BBH system in a Keplerian orbit, using a simple prescription to boost the BHs. Since we want to describe an inspiraling BBH using PN trajectories, we must use a coordinate system compatible with the PN gauge. As we explain in the next subsection, we use the PN trajectories in the standard harmonic coordinate system \cite{Blanchet:2013haa} and, for this reason, we build our superposition directly in harmonic coordinates \footnote{It was shown recently \cite{varma2018} that building superposed free initial data in harmonic coordinates, rather than in the standard Kerr-Schild coordinates, is more accurate for numerical simulations of BBH mergers. We also found that our superposed metric in harmonic coordinates is more accurate than the KS gauge.}.

The Kerr metric has a well-known harmonic coordinate system introduced by Cook and Scheel \cite{cook1997}, which is also horizon penetrating. This is an important feature of the coordinates for doing GRMHD simulations because the excision can be placed inside the horizon (we analyze more features of this harmonic coordinate system in Appendix A) . Using the known transformation from in-going Kerr to this harmonic coordinate system (c.f. Appendix B in Ref. \cite{ireland16}), we express the transformation from Kerr-Schild coordinates, $\lbrace x^a_{\rm KS} \rbrace$, to Cook-Scheel harmonic coordinates, $\lbrace x^a_{\rm H} \rbrace$, as:
\begin{align}
    t_{\rm KS} =& t_{\rm H} + 2M \log{(r-r_{-})} -2M \log(2M),
\\
    x_{\rm KS} =& x_{\rm H}+ M \left[ \frac{(r-M) y_{\rm H} - a x_{\rm H}}{(r-M)^2+a^2} \right],
\\  
    y_{\rm KS} =& y_{\rm H}+ M \left[ \frac{(r-M) x_{\rm H} + a y_{\rm H}}{(r-M)^2+a^2} \right],
\\ 
    z_{\rm KS} =& z_{\rm H}+ M \left( \frac{z_{\rm H}}{r-M} \right),
\end{align}
where we have:
\begin{equation}
    r-M = \sqrt{(Q+W)/2}, \quad W:= \sqrt{Q^2+4a^2 z_{\rm H}^2},
\end{equation}
and
\begin{equation}
    Q:= r_{\rm H}^2-a^2, \quad r_{\rm H}^2:= x_{\rm H}^2+y_{\rm H}^2+z_{\rm H}^2.
\end{equation}

The space components are thus related by the elegant relation:
\begin{equation}
    (x^i_{\rm KS}-x^i_{\rm H})\delta_{ij}(x^j_{\rm KS}-x^j_{\rm H}) = M^2.
\end{equation}

If we apply this transformation to the Cartesian Minkowski part of the Kerr-Schild metric \eqref{eq: ks}:
\begin{equation}
    \eta^{\rm H}_{ab}(a,M) = \frac{\partial x_{\rm KS}^{a'}}{\partial x_{\rm H}^{a}} \frac{\partial x_{\rm KS}^{b'}}{\partial x_{\rm H}^{b}} \eta_{a'b'},
\end{equation}
we note that the transformed quantity $\eta^H_{ab}(a,M)$ now depends on the spin and mass of the BHs. However, we can still write this as a flat Cartesian metric plus a \textit{source} term:
\begin{equation}
\eta^{\rm H}_{ab}(a,M) = \eta_{ab} + M \mathcal{A}_{ab}(a,M),
\end{equation} where $\eta_{ab}$ is again the Cartesian Minkowski metric. It can be shown that $\mathcal{A}_{ab}(a,M)$ is well-behaved at spatial infinity:
\begin{equation}
    \mathcal{A} \sim 1/r, \quad  \text{for }r\rightarrow \infty.
\end{equation}

The second term of the Kerr-Schild metric can be transformed in the same manner and is also well behaved at infinity. We conclude that the Kerr metric in harmonic coordinates can be written as a background plus a BH term, suitable for superposition, as:
\begin{equation}
    g_{ab}= \eta_{ab} + M \mathcal{H}_{ab},
    \label{eq: harmonicmetric}
\end{equation}
where $\mathcal{H}_{ab}:= 2 H l^{\rm H}_{a}l^{\rm H}_{b} + \mathcal{A}_{ab}$. In these coordinates, we shall build an effective metric (superposed harmonic PN, or SHPN) of the form:
\begin{equation}
    g_{ab}= \eta_{ab} + \phi^*_{(1)} \Big ( M_{(1)} \mathcal{H}^{(1)}_{ab} \Big) +
    \phi^*_{(2)} \Big ( M_{(2)} \mathcal{H}^{(2)}_{ab} \Big),
    \label{eq: BBHharmonicmetric}
\end{equation}
where $M_{(1)}$ and $M_{(2)}$ are the masses of each BH, $\mathcal{H}^{(1)}_{ab}$ and $\mathcal{H}^{(2)}_{ab}$ their corresponding tensorial functions $\mathcal{H}_{ab}$, and $\phi^*_{(1)}$ and $\phi^*_{(2)}$ are transformations that \textit{boost} the BH terms to describe the  global metric of a BBH system, using a PN approximation for the trajectories. We show how to build this transformation in the next sub-section. 

\subsection{Moving superposed black holes with PN trajectories}

The Kerr metric in harmonic coordinates \eqref{eq: harmonicmetric} represents a BH \textit{at rest} with respect to an asymptotically inertial frame. To describe a uniformly moving BH, we can apply a Lorentz boost transformation and convert our coordinates to boosted coordinates (c.f. Ref. \cite{penna2015}). Physical quantities at spatial infinity transform as four-vectors in Minkowski space-time. For instance, the asymptotic observer will measure that a boosted BH has a mass $M_{B} = \gamma M$, where $M$ is its rest ADM mass. This is simply a frame transformation, and it does not change gauge-invariant quantities such as the Ricci or Kretschmann scalars.

Let us suppose that we have a binary system, where the BHs move in an inspiraling orbit with respect to the origin of a (Cartesian) coordinate system $\mathcal{O}$, with their trajectories given by:
\begin{equation}
    s^a_A(t)=(t, \vec{s}_A(t)) \equiv (\gamma_A \tau_A, \vec{s}_A(\tau_A)),
    \label{eq-worldline}
\end{equation}
where $A\in\{$ BH$_1$, BH$_2\}$, $\vec{s}_A(t)$ is the spatial trajectory of a BH, $\tau_A$ the proper time, and $\gamma_A$ the Lorentz factor. Throughout this work, we assume that the BH spins are (anti-)aligned, which implies that there is no precession and the orbit lies on the $xy$ plane. 

For our approximate BBH metric, we are going to \textit{boost} two BH terms $\mathcal{H}_{ab}$ from Eq. \eqref{eq: harmonicmetric} and superpose them, as sketched in Eq. \eqref{eq: BBHharmonicmetric}. We build the time-dependent boosts as coordinate transformations from the BH frames $\mathcal{O}' \lbrace X^{a} \rbrace$, to the (global) center of mass frame $\mathcal{O}\lbrace x^{a} \rbrace$. In $\lbrace X^{a} \rbrace$, the BH is at rest and its metric is locally given by Eq. \eqref{eq: harmonicmetric}; in the global coordinates $\lbrace x^{a} \rbrace$, on the other hand, the BH is moving according to the worldline \eqref{eq-worldline}. This transformation constitutes a \textit{generalized} boost since the BHs are not in uniform motion, i.e., the BH coordinates $\lbrace X^{a} \rbrace$ are non-inertial coordinates. 

\begin{figure}[ht]
  \centering
  \includegraphics[width=1\linewidth]{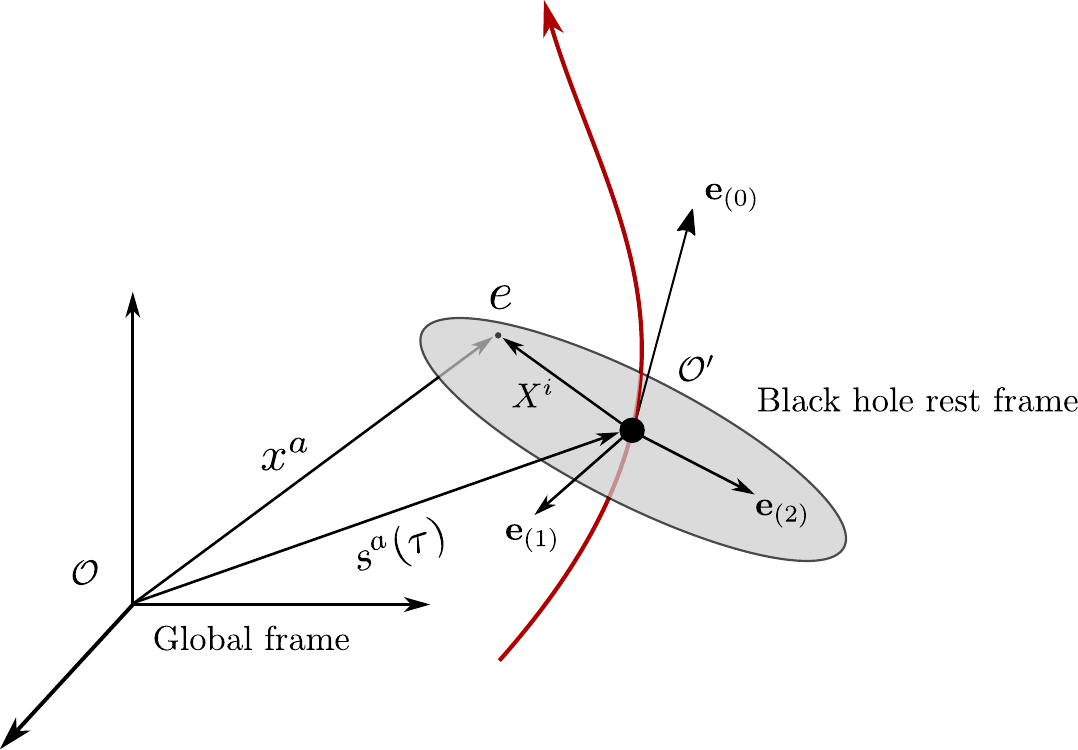}
  \caption{Diagram of the BH rest frame, $\mathcal{O}^\prime$, which is non-inertial, and the global frame, $\mathcal{O}$. The coordinates $\lbrace X^a \rbrace$ and $\lbrace x^a \rbrace$ both describe the event $e$.}
  \label{fig:frame}
\end{figure}

The natural (pseudo-Cartesian) coordinate system associated with the frame of an accelerated worldline is called a Fermi Normal coordinate system \cite{manasse1963fermi,poisson2011motion}. These widely used coordinates generalize the boost transformation for time-dependent velocities (see Ref. \cite{mashhoon2002length} for details). Let us say we want to build this coordinate transformation for a given event $e$ in spacetime (see Figure \ref{fig:frame}). First, find the proper time for which $s^{a}(\tau)$ is simultaneous to $e$ in the non-inertial BH frame. Then define the time coordinate of the system $\mathcal{O'}$ to be the proper time of the worldline $X^{0}=T=\tau$. Finally, assume that the hypersurface orthogonal to the worldline is approximately Euclidean, so the event $e$ described in the global coordinate system $\mathcal{O}$ is connected with $X^a$ as:
\begin{equation}
    x^{a} = s^{a}(\tau) + X^{i}e ^{a}_{(i)}(\tau),
    \label{eq:coordinates}
\end{equation}
where $e^a_{(i)}$ are the components of the orthonormal basis carried by the BH, in global coordinates.

To find the coordinate transformation we need to find the components of the orthonormal basis of the BH in the global coordinates. Let us assume that the frame carried by the BH is parallel to the axes of the inertial system $\mathcal{O}$. Since at each point the BH has a time-dependent velocity with respect to $\mathcal{O}$, locally we have to boost at each point in time the spatial frame to compare this with the global frame. The general Lorentz transformation in the $xy$ plane, given the spatial velocity $\vec{v} = \beta \vec{n}(t)$ of the BH, can be obtained with the boost generators, $\vec{K}$, and rapidity, $\xi= \tanh^{-1}(\beta)$, as:
\begin{align*}
\Lambda(t) &= \exp{(\xi \vec{n}(t) \cdot \vec{K})} \cr
& =
  \begin{pmatrix}
    \gamma & \gamma \beta n_{x} & \gamma \beta n_y & 0 \\
    \gamma \beta n_{x} & 1+ (\gamma-1) n_{x}^2 & (\gamma-1) n_x n_y & 0 \\
    \gamma \beta n_y & (\gamma-1) n_x n_y &  1+ (\gamma-1) n_{y}^2 & 0 \\
    0 & 0 & 0 & 1
  \end{pmatrix}.
\end{align*}

We use this transformation to boost the spatial basis of the BH. In the BH coordinates, this basis is simply given by $e^{a'}_{(i)} = \delta^{a'}_{i}$ (i.e. the Cartesian spatial basis). Then, in global coordinates this is given by $e^{a}_{(i)}= \Lambda ^{a}_{i}(t)$.

The coordinate transformation, using (\ref{eq:coordinates}), is given by:
\begin{align}
    t  &= \gamma \left[ T - \beta ( n_x X + n_y Y) \right] ,\\
    x &= s_x(t) +  X \left[1+ (\gamma-1)n_x^2\right]
         +Y\left[ -(\gamma-1) n_x n_y)\right], \\
    y &= s_y(t) + X \left[-(\gamma-1) n_x n_y \right]
         + Y \left[1+ (\gamma-1)n_y^2\right] ,\\
    z & = Z.
\end{align}

The non-inertial coordinates $X^a$ in terms of the global coordinates $x^a$ are easily obtained inverting these equations. Note that this transformation reduces to a standard Lorentz boost if the trajectory of the BH is a straight line with uniform rapidity.  superposing the two BH terms and performing this transformation for each term, with worldlines $s^a_1(t)$ and $s^a_2(t)$, we have explicitly our SHPN metric:
\begin{equation}
\boxed{g_{ab} = \eta_{ab} + M_1  \Big (\frac{\partial X_1^{\overline{a}}}{\partial x^{a}}\frac{\partial X_1^{\overline{b}}}{\partial x^{b}} \mathcal{H}_{\overline{a} \overline{b}} \Big) + M_2 \Big (\frac{\partial X_2^{\overline{a}}}{\partial x^{a}}\frac{\partial X_2^{\overline{b}}}{\partial x^{b}}  \mathcal{H}_{\overline{a} \overline{b}} \Big),}
\label{eq-metric}
\end{equation}
where the tensors are transformed through the Jacobian of the coordinates $X^a_A(x)$. We still have to supplement the metric with the position, velocity, and acceleration of the BHs. In the case of a BBH, we can obtain those solving the PN equations of the system in harmonic coordinates, as we show in the next section.

\subsection{Post-Newtonian trajectories for spinning BH binaries}

We assume for now that the orbit of the binary has circularized and the system is well described by the so-called adiabatic approximation \cite{Blanchet:2013haa}. We also assume that the spins of the holes are (anti-)aligned with the orbit and thus we ignore orbital precession. In this case, the inspiral is driven by the loss of binding energy of the orbit, $E$, balanced by the gravitational wave flux of energy, $\mathcal{F}$, and change in mass $\dot{M}$:
\begin{equation}
\dot{E} = -\mathcal{F} -\dot{M},
\label{eq: balance}
\end{equation}
where a dot represents a derivative with respect to the global time $t$. From this equation, we can obtain the orbital phase, $\Phi(t)$, and separation, $r_{12}(t)$, of the system. In the case of quasi-circular orbits, the gauge-dependent separation $r_{12}$ is linked to the orbital frequency through the relativistic generalization of Kepler's law \cite{Blanchet:2013haa}. 

First, we solve for the orbital phase. We rewrite Eq. \eqref{eq: balance} as two equations in terms of the (gauge invariant) variable $v:= (M d\Phi/dt)^{1/3}$:
\begin{align}
\frac{dv}{dt} &= - \frac{\mathcal{F}(v) + \dot{M}(v)}{dE(v)/dv},
\label{eq-taylort4}
\\
\frac{d\Phi}{dt} &= \frac{v^3}{M},
\label{eq-phaset4}
\end{align}
and replace $E(v)$, $\mathcal{F}(v)$, and $\dot{M}(v)$, with their explicit expressions at 3.5 PN order for the case of non-precessing binaries, as presented in
Ref.~\cite{Ajith:2012tt}.

Following the TaylorT4 scheme~\cite{Buonanno:2002fy}, we expand the right hand side of Eq.~\eqref{eq-taylort4} in a Taylor series to the proper PN order, and integrate it to obtain $t(v)$ (see, e.g., Refs.~~\cite{Damour:2000zb,Damour:2002kr,Buonanno:2009zt} for Taylor 
PN approximants). Then, we invert this quantity, and solve
Eq.~\eqref{eq-phaset4} for $\Phi(t)$. Next, we solve for the orbital separation $r_{12}(t)$ within numerical accuracy although it is directly derived from the orbital frequency in the PN approximation. We write its time derivative in terms of the orbital energy:
\begin{equation}
\dot{r}_{12} = \frac{dE/dt}{dE/dr_{12}} \equiv -\frac{ \mathcal{F}(r_{12}) + \dot{M}(r_{12}) }{dE/dr_{12}},
\end{equation}
and integrate to find:
\begin{equation}
\label{eq: t_r12}
t(r_{12}) = t_c - \int^{r_{12}}_{0} d\tilde{r}_{12} \frac{dE/d\tilde{r}_{12}}{ \mathcal{F}(\tilde{r}_{12}) + \dot{M}(\tilde{r}_{12}) } \,,
\end{equation}
where $t_c$ is the time until coalescence. We replace $E(r_{12})$, $\mathcal{F}(r_{12})$, and $\dot{M}(r_{12})$ with their explicit expressions, found in Ref. \cite{Blanchet:2013haa} and references therein, and solve for  $t(r_{12})$. Finally, we invert using a Newton-Rawson method to obtain $r_{12}(t)$. With $\Phi(t)$ and $r_{12}(t)$ we can reconstruct the worldlines \eqref{eq-worldline} of each spinning hole in harmonic coordinates, as required  by our metric \eqref{eq-metric}. 

Note that, even though we use the PN approximation to obtain the BH trajectories, our metric is valid in the inner zone because we are using the full relativistic BH terms that include the ergosphere and horizon (see Appendix A). In other words, we are restricted to binaries with separations larger than $\sim 10M$, but the metric, as we will show now, is accurate in both inner and near zones.

\section{Analysis of the spacetime metric}

In this section, we test the global validity of our SHPN metric \eqref{eq-metric}. The metric of a BBH system must satisfy Einstein's field equation in a vacuum, and thus the Ricci tensor must be zero. In numerical relativity,  violations of Einstein's equation are tracked using the Hamiltonian and momentum constraints. Since we intend to use the four-dimensional form of the metric in our applications, here we focus first on four-dimensional quantities to quantify deviations from the exact solution. In particular, following Ref. \cite{Mundim:2013vca, ireland16}, we investigate the Ricci scalar, $R:=g^{ab}R_{ab}$, where $R_{ab}$ is the Ricci tensor. Violations of the Ricci scalar $R$ are not absolute and, thus, they are only meaningful when compared with other quantities. For instance, if we have that $|R(t_1)|>|R(t_0)|$ for $t_1>t_0$, we can state that the approximate metric has deteriorated  or deviated from a vacuum solution over time. Similarly, we can compare the Ricci scalar of different systems or at different points in space to assess locally where the metric is a better approximation to a vacuum solution.

\begin{figure}[htbp]
  \centering
  \includegraphics[width=0.9\linewidth]{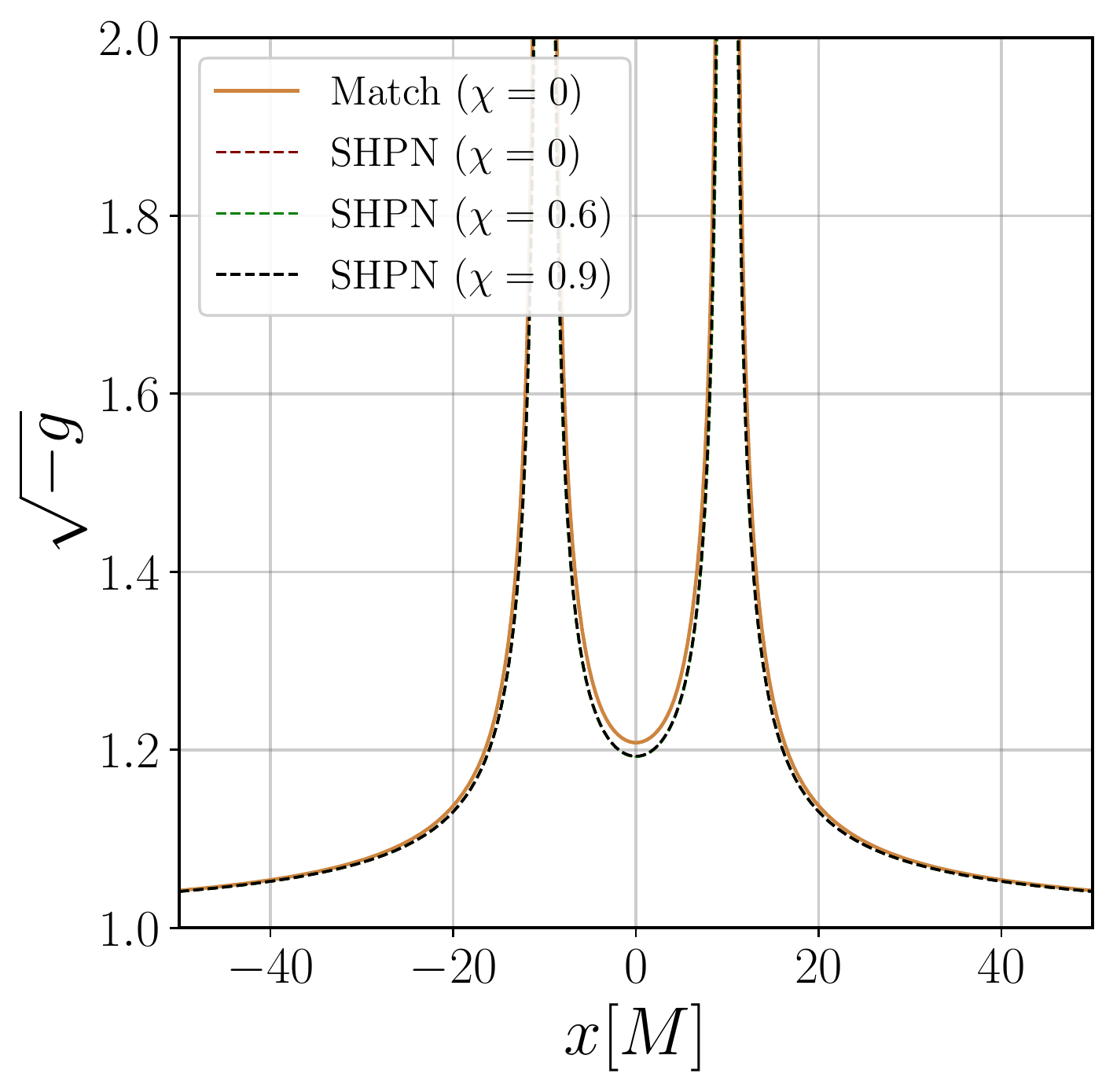}
  \caption{Determinant of superposed metric for different values of the spin and separation of $r_{12}=20M$. Note that the curves for different spins are very similar. For comparison, we include the determinant of the matching metric for $\chi=0$.}
    \label{fig: detg}
\end{figure}

\begin{figure*}[htbp]
  \begin{center}
  \includegraphics[width=.75\textwidth]{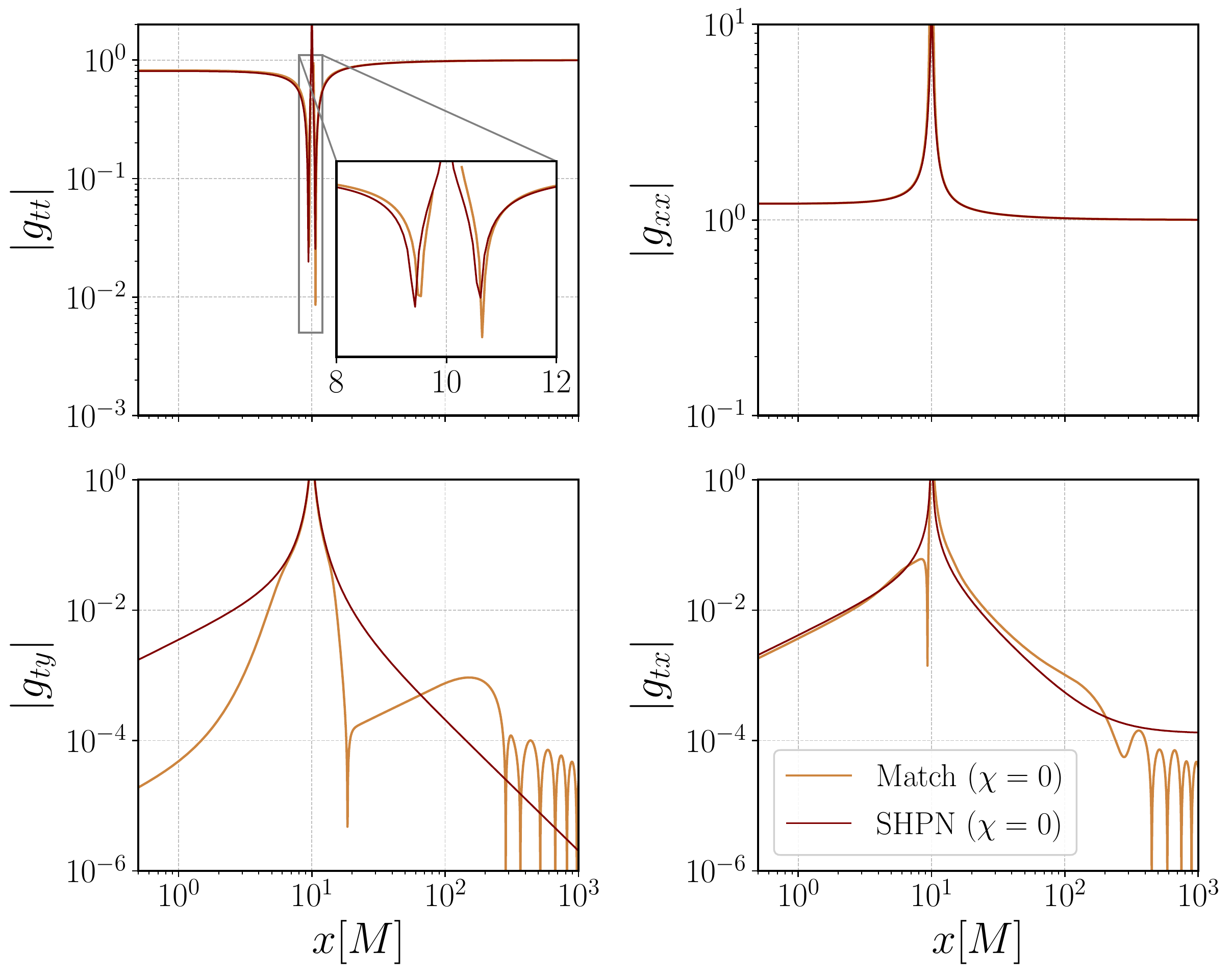}
  \end{center}
  \caption{Absolute value of several metric components for the superposed and matching metric in the fiducial configuration. Note that the superposed metric components are much smoother than the matching metric because there are no transition regions.}
  \label{fig: metric}
\end{figure*}

We also compare the validity of our solution with the alternative approach presented in Refs. \cite{Mundim:2013vca, ireland16}, where an analytical metric is built by stitching different approximated solutions of Einstein's equations at three characteristics zones of a binary compact system, namely, the Inner-Zone (IZ), the Near-Zone (NZ), and the Far-Zone (FZ)  \cite{Yunes:2006bb, Yunes:2006iw}. This so-called asymptotic matching procedure brings all these different parts into the same harmonic coordinate system and the global metric can be written as:
\begin{eqnarray}
 g_{ab} &=&
 (1 - f_{\rm FZ})
 \Bigl\{f_{\rm NZ} \bigl[f_{{\rm IZ},1} \,g_{ab}^{\rm (NZ)} 
 +(1 - f_{{\rm IZ},1} ) \,g_{ab}^{\rm (IZ1)}\bigr]
 \nonumber \\ &&
 + (1 - f_{\rm NZ} )\bigl[f_{{\rm IZ},2} \,g_{ab}^{\rm (NZ)} 
 +(1 - f_{{\rm IZ},2} ) \,g_{ab}^{\rm (IZ2)}\bigr]\Bigr\}
 \nonumber \\ &&
 + f_{\rm FZ} \,g_{ab}^{\rm (FZ)} \, ,
 \label{eq:wholemetric}
\end{eqnarray}
where transition functions $f_i$ are used to go from one zone to the other. This analytical metric, however, is computationally expensive and complex to handle. The Jacobians required to stitch the different parts of the metric into the same coordinate system are very long, and many operations are required to compute them at each timestep. Moreover, for the spinning case, the matching procedure renders the metric prohibitively expensive for MHD simulations. In our new approach, we lose some accuracy in comparison with the matching metric but we gain much more efficiency.

\subsection{Spacetime scalars}

Although the metric is analytical, we compute its spacetime scalars numerically as it is faster and more practical to incorporate the PN trajectories. The functional form of the metric is built using \texttt{Mathematica} \cite{mathematica} and exported to C language in an optimized form. We use then a C-based code that implements fourth-order finite differences in a Cartesian grid for the derivatives of every metric function. We analyze and plot the outputs using \texttt{Numpy} and \texttt{Matplotlib} \cite{numpy, matplotlib} The convergence analysis of these methods is presented in Appendix B.

We are interested in using the metric in the inspiraling regime, where the PN approximation holds, and the system is emitting a significant amount of gravitational radiation. We explore the characteristics of the system for a fiducial configuration, with a separation of $r_{12}(t_0) =20 M$, equal BH masses, and the adimensional spin parameter, $\chi:=a/M$, in the interval $0 < \chi < 0.9$.

As a first check of consistency, we analyze the metric determinant $\sqrt{-g}$. In Figure \ref{fig: detg}, we plot the determinant for a separation of $r_{12}=20M$ and several values of the spin parameter $\chi$, along with the determinant for the matching metric. We see that for all these values, the determinant for the superposed metric is globally well-behaved, free of pathologies, and similar to the matching space-time.

In Figure \ref{fig: metric}, we plot some components of both metrics. It is interesting to note that the $g_{tt}$ component of the SHPN metric is globally similar to the matching one, meaning that the effective PN potential of both spacetimes is much akin \cite{bowen2017}. The differences between the two metrics are important in the transition regions and the Far-Zone. In the latter, the matching metric incorporates the post-Minkowski background of gravitational waves, while our new SHPN is asymptotically flat; however, we do take into account the gravitational radiation losses in the trajectories of the BHs.

\begin{figure}[htb]
  \centering
  \includegraphics[width=1\linewidth]{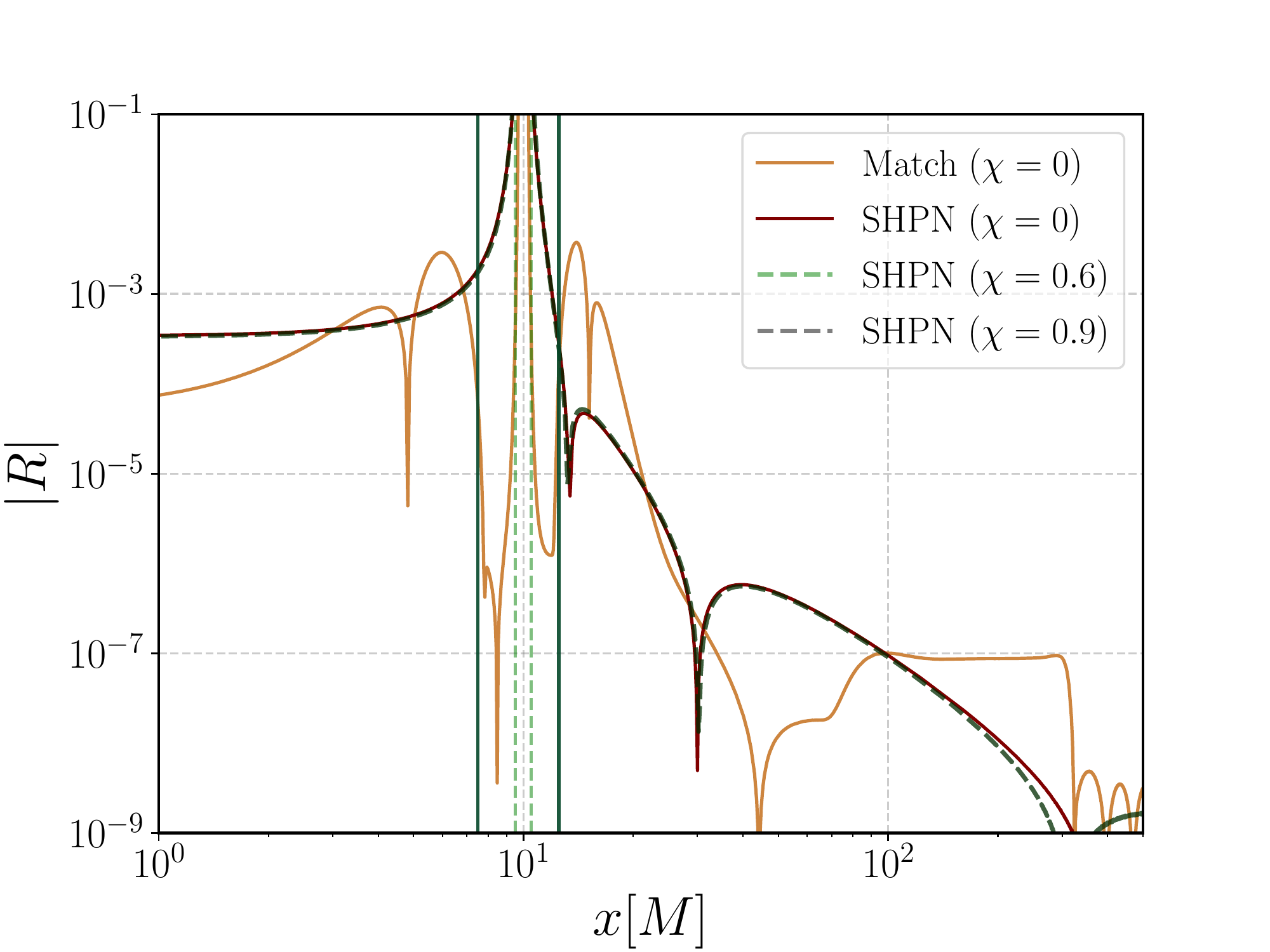}
  \caption{Ricci scalar for the matching and superposed metric with $r_{12}=20M$, equal mass-ratio, and different values of spin. The dashed green lines denote the location of the horizon and the solid  green lines the location of the ISCO for a non spinning BH.}
  \label{fig: riccicomparison}
\end{figure}

\begin{figure}[htbp]
  \centering
  \includegraphics[width=1\linewidth]{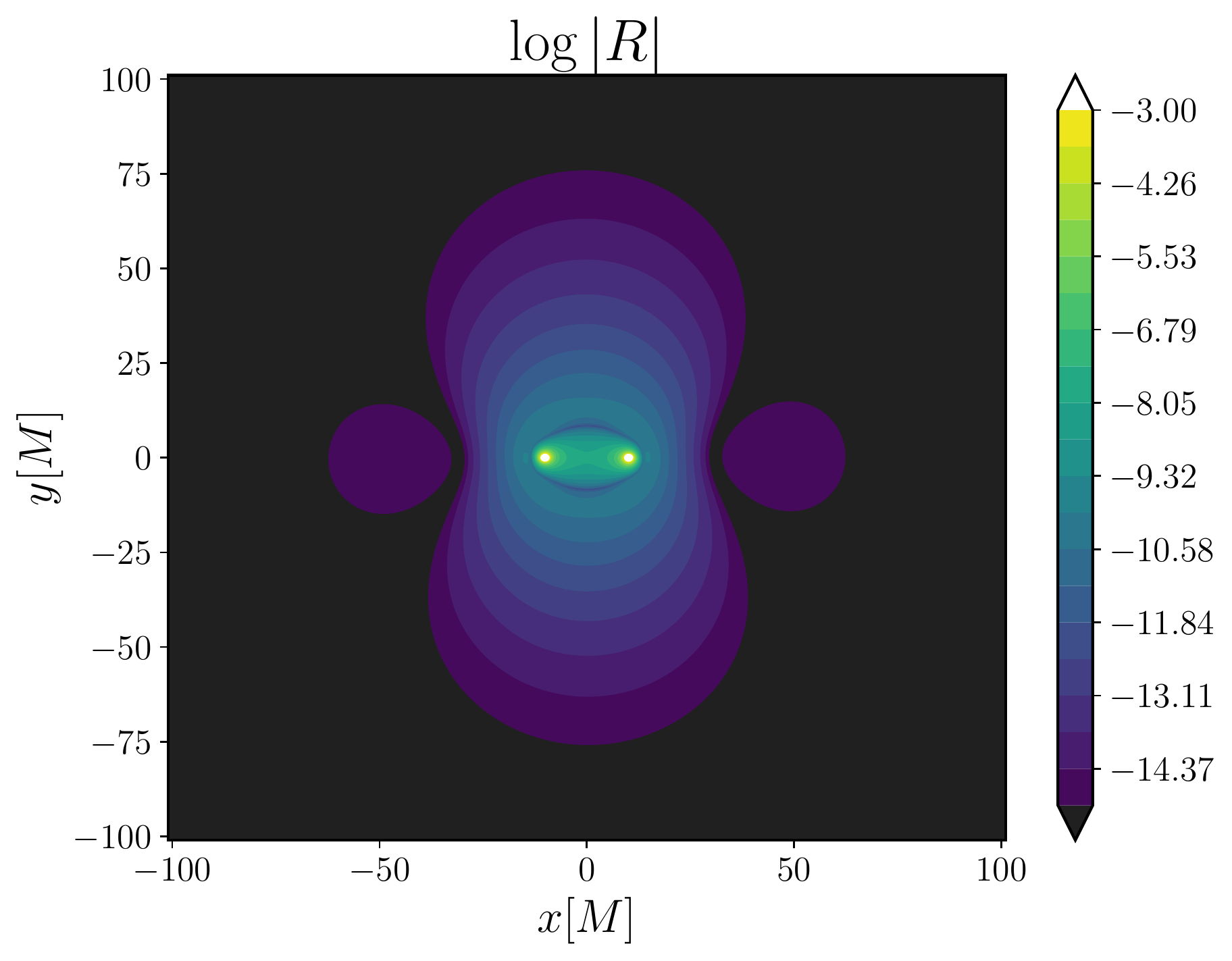}
  \caption{Equatorial logarithmic plot of the Ricci scalar for the SHPN metric with $r_{12}=20M$, equal masses, and $\chi=0.9$}
  \label{fig:2dricc}
\end{figure}

\begin{figure}[htbp]
  \centering
  \includegraphics[width=1\linewidth]{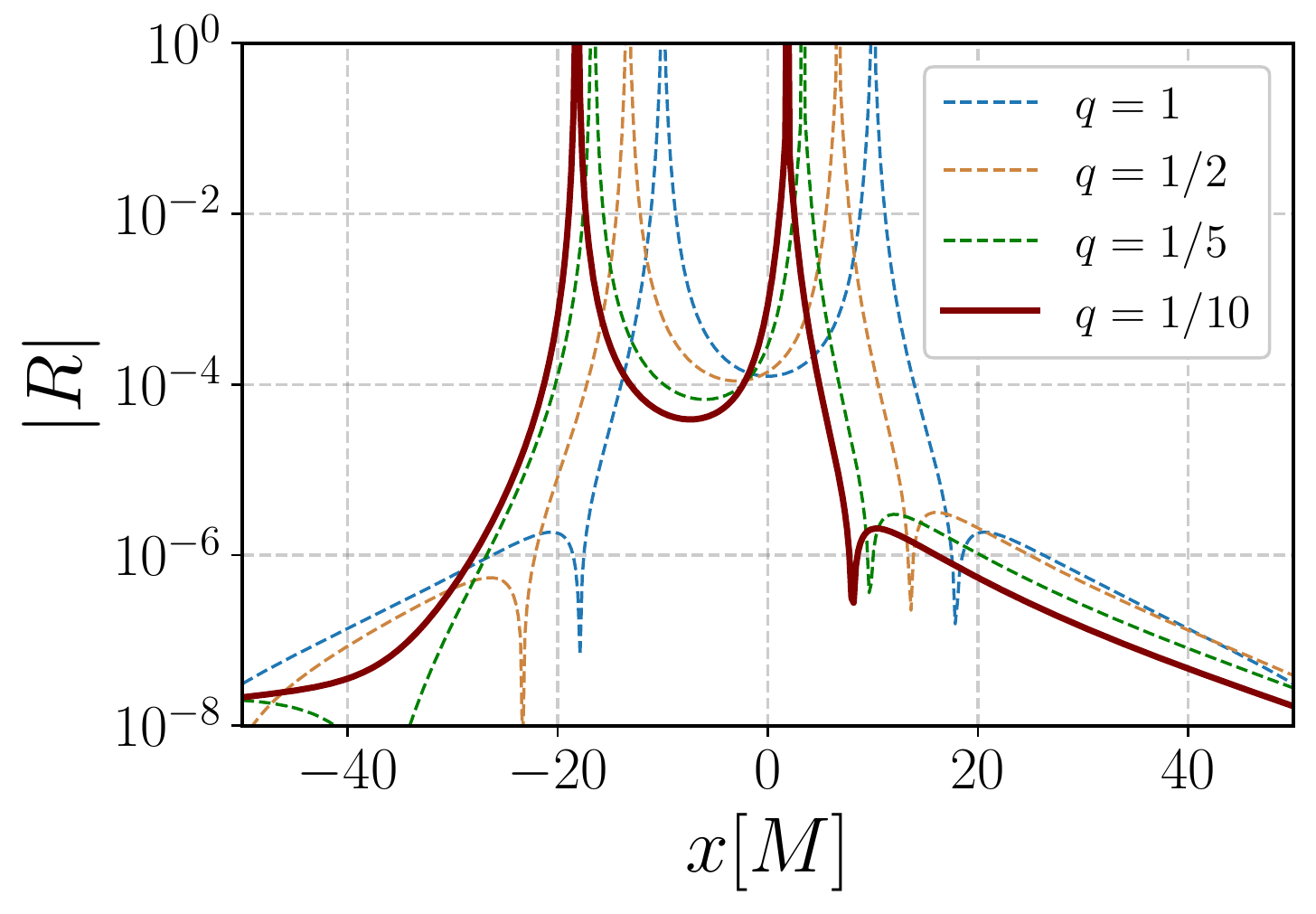}
  \caption{Ricci violations of the SHPN metric, for $r_{12}=20M$, $\chi=0.5$, and different mass-ratio values. }
  \label{fig-massratio}
\end{figure}

In Figure \ref{fig: riccicomparison}, we plot the Ricci scalar of the SHPN metric over the positive $x$ axis at $z=y=0$, for different values of spin, and we compare it with the Ricci scalar of the matching metric for a binary of the same characteristics. First, we see that $R$ varies very little under different spin parameters, consistent with Ref. \cite{ireland16}. Note that the matching metric is better in the IZ but the violations are worse at the transition regions outside the ISCO, where the SHPN is smoother and performs better. A good metric accuracy in this region is an important feature for determining the correct gas dynamics of an accreting disk near the hole. In Figure \ref{fig:2dricc}, we show an equatorial plot of the Ricci scalar. As expected, the higher violations are concentrated in the middle region between the BHs and drop sharply with distance. In Figure \ref{fig-massratio}, we plot the Ricci violations for different mass-ratios $q:=M_1/M_2$. We find the values of $R$ depend smoothly on $q$, improving in the middle region for smaller $q$.

Besides $R$, we can explore other curvature scalars to assess the global behavior of the metric. In particular, considering the ADM equations for a general spacetime, we can define the Hamiltonian constraint $\mathcal{H}$ as:
\begin{equation}
\mathcal{H}:= {}^{3}R + K - K_{ab} K^{ab} = 16 \pi \tilde{\rho},
\end{equation}
where ${}^{3}R$ is the spatial Ricci curvature, $K_{ab}$ the extrinsic curvature, and $\tilde{\rho}$ the energy density of matter. For our BBH vacuum metric, a non-zero $\mathcal{H}$ means that the spacetime has ``fake mass'' due to the approximation. This will introduce errors in the true gravitational potential and thus in the geodesic motion of matter. Since we are interested in using this spacetime as a background scenario for evolving an MHD fluid, it is important to analyze this quantity and its evolution. We consider the volume-integrated value of $\mathcal{H}$ as a measure of the total fake mass introduced by the approximated metric:
\begin{equation}
M_{\rm fake} = \frac{1}{16 \pi}\int_{\mathcal{V}} \mathcal{H} \: d\mathcal{V}.
\end{equation}

\begin{figure}[ht!]
  \centering
  \includegraphics[width=1\linewidth]{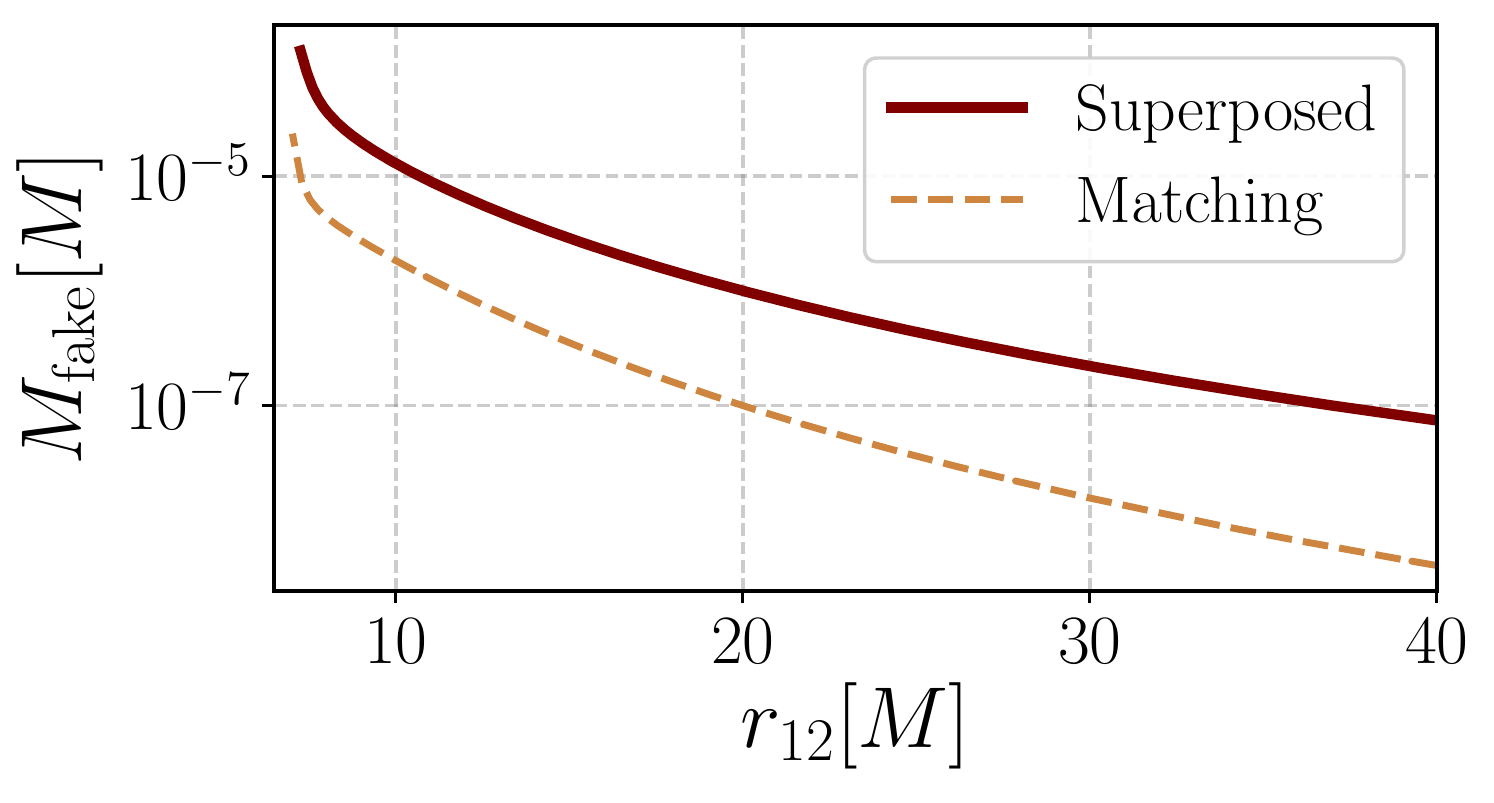}
  \caption{Fake mass introduced in the spacetime by the SHPN and matching metric approximations for different orbital distances. }
  \label{fig-fakemass}
\end{figure}

Considering a cube of radius $r=50M$ around the center of mass, we can track the evolution of $M_{\rm fake}$ for different orbital separations. As we show in Figure \ref{fig-fakemass}, this fake mass is overall small with respect to the total mass of the BBH in both SHPN and matching metric but starts increasing exponentially at $\sim 8M$, where the PN approximation breaks. 

\begin{figure}[ht!]
  \centering
  \includegraphics[width=1\linewidth]{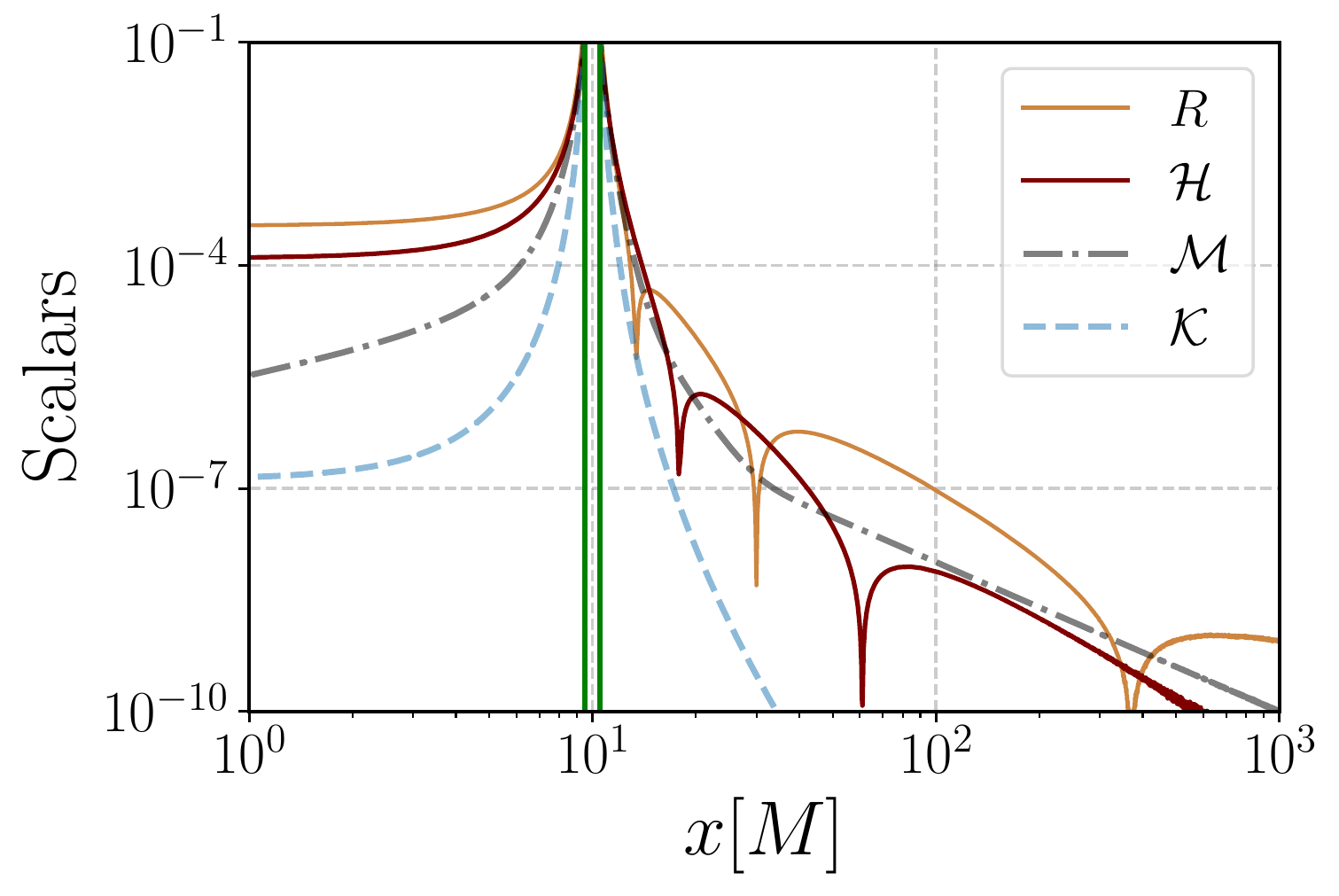}
  \caption{The Ricci scalar ($R$), the Hamiltonian ($\mathcal{H}$) and momentum ($\mathcal{M}$) constraint, and the Kretschmann scalar $\mathcal{K}$ of the SHPN metric, for a BBH with $r_{12}=20M$, equal mass, and $\chi=0.9$. We can notice that the Kretschmann scalar follows the decay $\sim 1/r^6$ typical for a single BH. The solid  green lines denotes the location of the ISCO for a non-spinning BH. }
  \label{fig-scalars}
\end{figure}

Finally, in Figure \ref{fig-scalars}, we plot the Ricci scalar, the Hamiltonian constraint, the square root of the Momentum constraints, $\mathcal{M}$, and the Kretschmann scalar, $\mathcal{K}:= R_{abcd}R^{abcd}$,  for the SHPN metric. We observe here that the Hamiltonian constraint and the Ricci scalar have similar behaviors, indicating that the errors of the approximation come essentially from the fake mass component.

\section{Superposed metric in GRMHD simulations}

\begin{figure*}[htbp]
  \begin{center}
  \includegraphics[width=.45\linewidth]{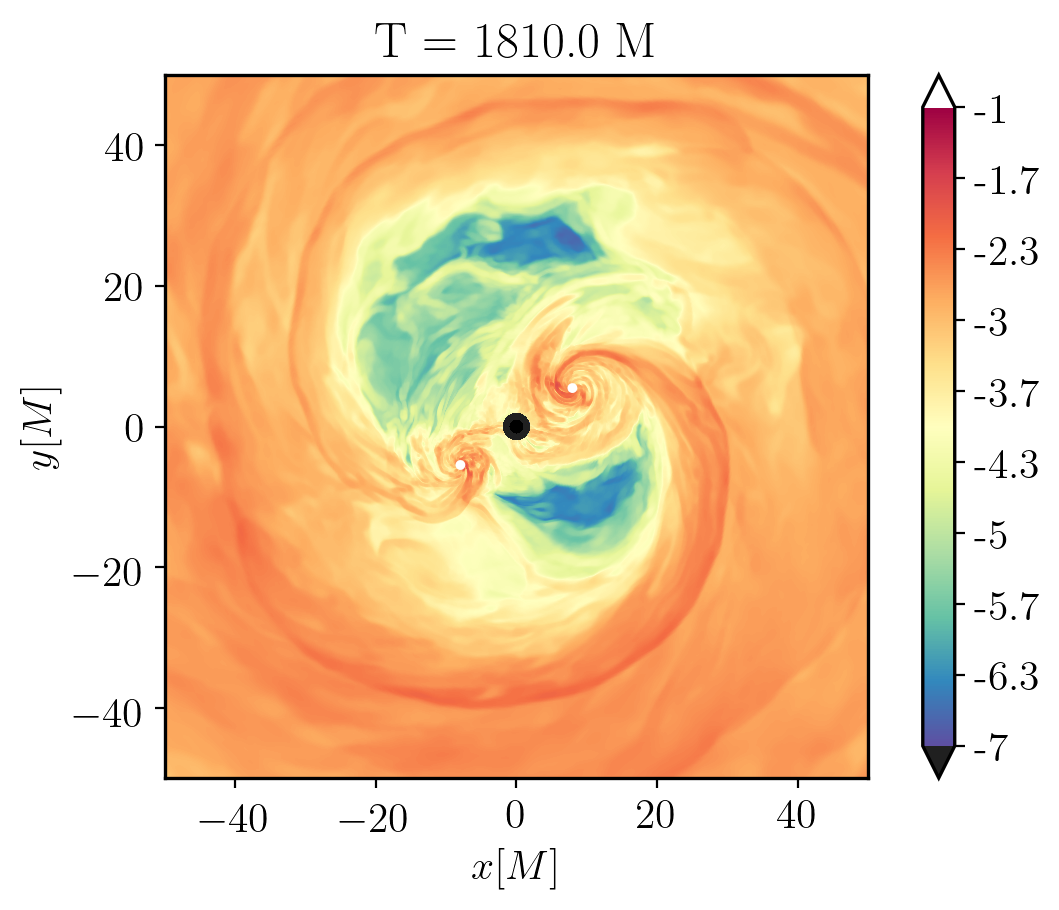}
  \includegraphics[width=.45\linewidth]{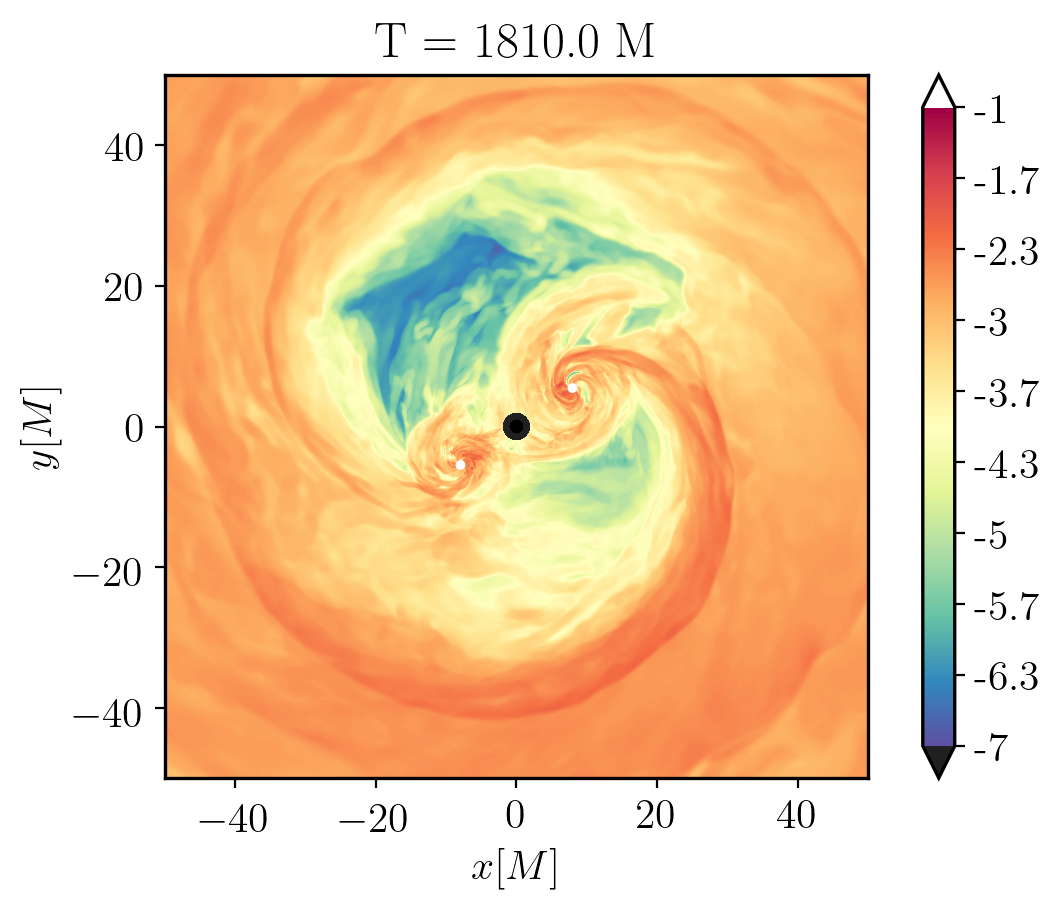} 
  \end{center}
  \caption{Equatorial plot of the rest-mass density $\rho$ in logarithmic scale for (right) the superposed metric and (left) the matching metric, at $T=1810M$, which represents approximately $\sim 3$ orbits.}
  \label{fig: densitysim}
\end{figure*}

We built our new superposed metric in harmonic coordinates, which allows us to use accurate PN trajectories and directly compare our simulations with previous results that use the same gauge. Moreover, the metric is accurate enough near the BH, allowing us to analyze what happens with the plasma physics around each BH. 

The matching metric approach has been tested in simulations, both as a background spacetime for MHD \cite{bowen2018,bowen2019}, and in numerical relativity, \cite{sadiq2018}. MHD simulations of BBHs accretion disks with the spinning matching metric, as we said, are prohibitively expensive. In contrast, the SHPN metric is computationally cheaper, allowing us to simulate this type of system for the first time. In this section, we show the results of a full 3D GRMHD simulation with the SHPN metric using the code \harm. In particular, we focus on comparing an accretion disk simulation using the SHPN metric for non-spinning BHs and the analog simulation presented in Refs. \cite{Bowen17, Bowen18}, which uses the non-spinning matching metric. We will present full details of the spinning BBH simulation with the new metric in an upcoming work \cite{combi2021}.

\subsection{GRMHD evolution}

Assuming the surrounding gas does not influence the spacetime dynamics, we can use our superposed metric to simulate the MHD evolution of accretion disks in a BBH system. For that purpose, we implement the new metric in the GRMHD code \harm \citep{GMT03,Noble06,Noble09,Noble12}, which evolves the ideal GRMHD equations in flux-conservative form, for an arbitrary metric and coordinate system. The equations of motion are the continuity equation, the local
conservation of energy and momentum, and Maxwell's
equations, which can be written as:
\begin{equation}
\partial_t \bU\left(\prim\right) =                                                                       
-\partial_i \bF^i\left(\prim\right) + \mathbf{S}\left(\prim\right) \ ,
\label{cons-form-mhd}
\end{equation}
where $\bP$ are the \textit{primitive} variables, $\bU$ are the \textit{conserved} variables, $\bF^i$ the \textit{fluxes}, and $\mathbf{S}$ are the \textit{source} terms. These can be expressed as
\begin{eqnarray*}
\bP & := & \left[ \rho, u, \tilde{u}^i, B^i \right], \\
\bU\left(\prim\right) & := & \sqrt{-g} \left[ \rho u^t ,\, {T^t}_t +
  \rho u^t ,\, {T^t}_j ,\, B^k\right] \ , \label{cons-U-mhd} \\
\bF^i\left(\prim\right) & := & \sqrt{-g} \left[ \rho u^i ,\, {T^i}_t +
  \rho u^i ,\, {T^i}_j ,\, \left(b^i u^k - b^k u^i \right)\right] \ , \label{cons-flux-mhd} \\
\mathbf{S}\left(\prim\right) & := & \sqrt{-g} \left[ 0 ,\,
  {T^\kappa}_\lambda {\Gamma^\lambda}_{t \kappa} - \mathcal{F}_t ,\,
  {T^\kappa}_\lambda {\Gamma^\lambda}_{j \kappa} - \mathcal{F}_j ,\, 0
  \right] , \label{cons-source-mhd}
\end{eqnarray*}    
where ${\Gamma^c}_{ab}$ 
are the Christoffel symbols,  $b^{a} = \left(1/u^t\right)\left({\delta^a}_b + u^a u_b \right)B^b$ is the magnetic 4-vector projected into the fluid's comoving reference frame, $B^i$ is the magnetic field in the reference frame of the space normal hypersurface, $u$ is the internal energy density, $u^{\alpha}$ are the components of the fluid's 4-velocity, and $\tilde{u}^i$ is the fluid velocity in the zero-angular-momentum observer (ZAMO) frame. The stress-energy tensor is written as
\begin{equation}
  T_{ab} = \left( \rho h + 2p_m \right) u_{a} u_{b} + \left( p + p_m\right)g_{ab} - b_{a}b_{b} \ ,
\end{equation}
where $h = 1 + \epsilon + p/\rho$ is the specific enthalpy, $\epsilon:=u/\rho$ is the specific
internal energy, $p$ is the gas pressure, $p_m = \frac{1}{2}b^2$ is the magnetic pressure, and $\rho$ is the rest-mass density. We include a source term $\mathcal{F}_{\nu}$ into the local energy conservation equation in order to approximate effects from radiative cooling, designed to preserve the height ratio of the disk \cite{Bowen17}. We assume and ideal $\Gamma$-law equation of state: $p = (\Gamma-1) \rho \epsilon$, with $\Gamma = 5/3$. 

\harm uses high-resolution shock-capturing methods to integrate the conservation equations \eqref{cons-form-mhd}. In particular, we use the piece-wise parabolic reconstruction of primitive variables for the local Lax-Friedrichs fluxes, a Flux-CT scheme to maintain the solenoidal constraint \cite{toth2000}, and a robust recovery procedure from conserved to primitive variables \cite{noble2006}. The code uses fourth-order finite differences for spatial derivatives of the metric to find the Christoffel symbols, and the method of lines for time integration with a Runge-Kutta method of second-order (more details of the algorithm in Refs. \cite{Noble09} and \cite{Noble12}).

\subsection{Mini-disk dynamics in a binary black hole system: comparison with previous simulations}

In Ref. \cite{Bowen18}, the matching metric \eqref{eq:wholemetric} was used to simulate the MHD dynamics of a circumbinary disk around a BBH and the formation of mini-disks. Ref.  \cite{bowen2018} showed, for the first time, that mini-disks in tight binary systems are out of inflow equilibrium, filling and depleting their mass in less than an orbital period, showing interesting modulations \cite{bowen2019,dAscoli:2018fjw}. 

We perform a simulation with the same configuration and initial data as Refs. \cite{bowen2018,bowen2019} but switching the matching metric for our superposed metric \eqref{eq:wholemetric} with zero spin. The simulation uses a double-fisheye spherical grid \cite{zilhao2014} that focuses more cells in the vicinity of the BHs and maintains a spherical topology at the circumbinary region. The initial separation of the BHs is $r_{12}=20\:M$ and the initial data for the matter fields are constructed from a stabilized snapshot of the circumbinary simulation performed in Ref. \cite{Noble12}, with additional quasi-equilibrated mini-disks around each BH (see Ref. \cite{bowen2019} for more details). Since we are starting with the same initial data and grid, we re-normalize the primitive $B^i$ field by $\sqrt{-g_{\rm match}}/\sqrt{-g_{\rm sup}}$ to maintain the solenoidal constraint of the field after switching to the new metric.

\begin{figure}[htbp]
  \begin{center}
  \includegraphics[width=1\linewidth]{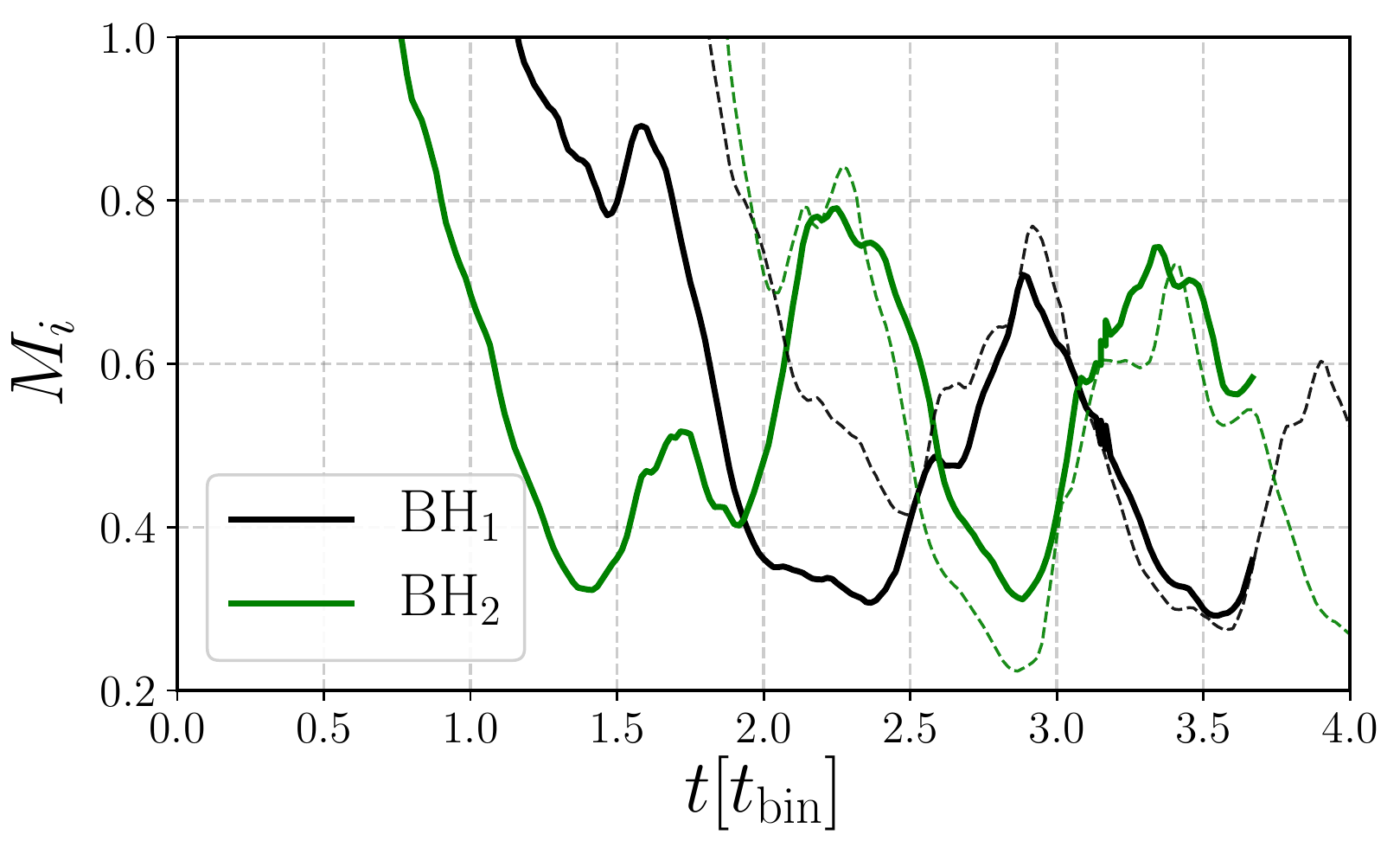}
  \end{center}
 \caption{Mass in each BH's mini-disk region for the superposed metric simulation (thick lines) and the matching metric simulation (dashed lines).}
    \label{fig: mass}
\end{figure}

\begin{figure}[htbp]
  \begin{center}
  \includegraphics[width=1\linewidth]{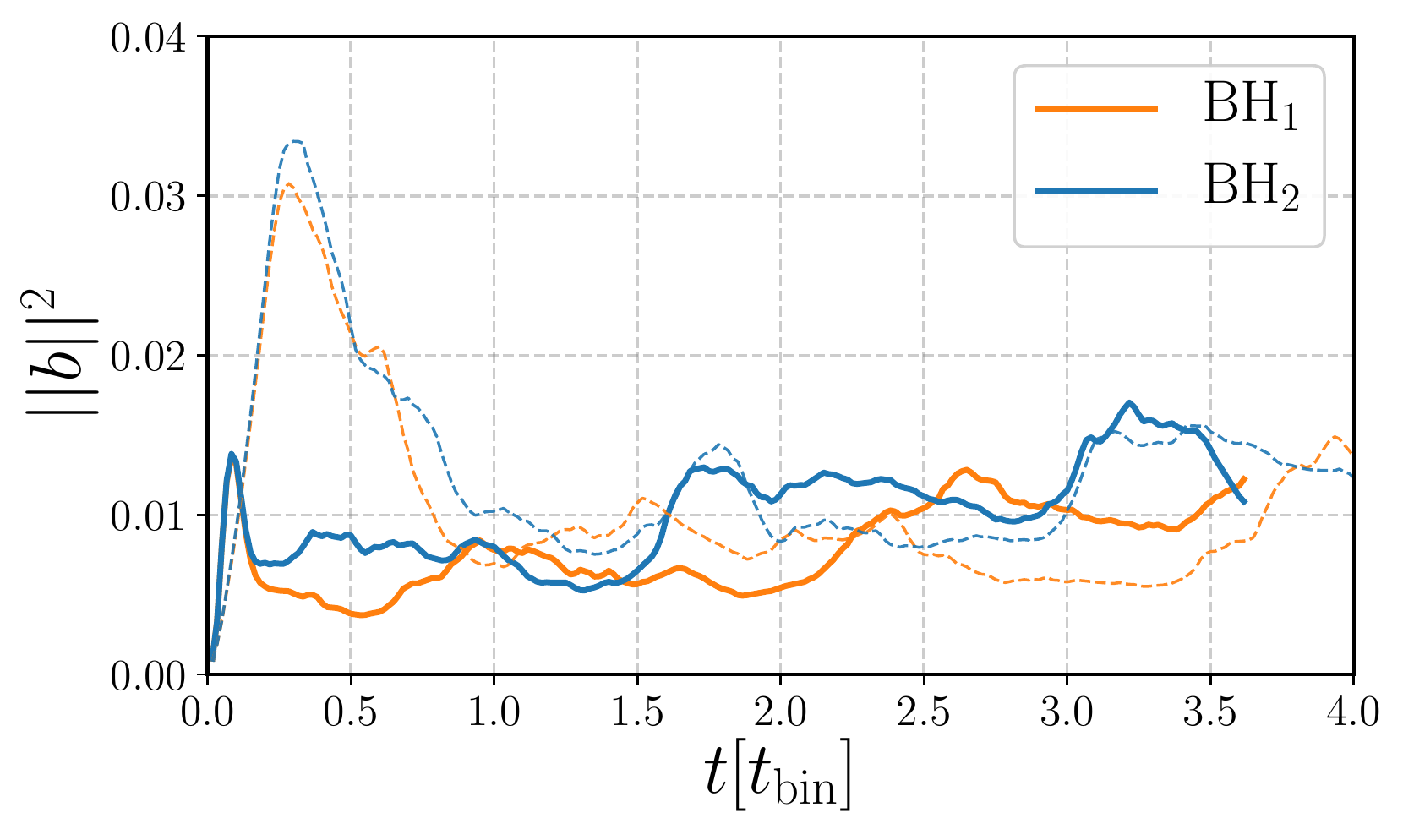}
  \end{center}
  \caption{Magnetic energy in each BH's mini-disk region for the superposed metric simulation (thick lines) and the matching metric simulation (dashed lines).}
    \label{fig-bsq}
\end{figure}

As shown in Ref. \cite{bowen2019}, the simulation has an initial transient that lasts for $\sim 2$ orbits. We thus evolve the system with the new metric for $\sim 3.5$ orbits as was done in \cite{bowen2018}. After equilibration, both matching and superposed metric simulations are in good agreement, as it can be seen, for instance, from the equatorial density snapshots in Figure \ref{fig: densitysim}. For a more quantitative assessment of both simulations we analyze the evolution of the mass contained in each mini-disk (Figure~\ref{fig: mass}), defined as the integrated rest-mass density:
\begin{equation}
M_i = \int_{\mathcal{V}_i} \rho u^0 \: \sqrt{-g} d^3x,
\end{equation}
where we take $\mathcal{V}_i$ as a spherical volume between the BH horizon, $r_i=r_{\mathcal{H}}$, and $r_f=0.4a(t)$, which is close to the Newtonian truncation radius \cite{bowen2017}. We also investigate the volume integral of the magnetic energy $b^2$ of each mini-disk (see Figure~\ref{fig-bsq}).

As we mentioned, for an equal-mass binary, the mini-disks are subject to a filling and depletion cycle. While the circumbinary lump feeds material to one of the mini-disks, the plasma in the other BH is completely accreted, and the BH \textit{starves}. After the initial transient, we see a remarkable overlap of each mini-disk mass $M_i$ for both simulations. The cycle is evident earlier in the superposed metric simulation. Since we are using the same initial data for both simulations (not just the same prescription), the equilibration of the mini-disks changes in the transient phase for the superposed, as the equilibrated mini-tori were set up using the matching metric. We also observe a good agreement in the behavior of the magnetic energy contained in each mini-disk after the transient.

\section{Conclusions}

We have presented a new approximate solution of Einstein's field equations for a spinning BBH in the inspiral regime. We built this solution as a linear superposition of boosted Kerr BHs in harmonic coordinates, supplemented with PN trajectories at 3.5 PN order. We compared our new metric with the well-tested asymptotic matching approach through an analysis of spacetime scalars. Although the matching approach has better accuracy in some specific regions, we found that the superposed metric has comparable accuracy, is smoother, and much cheaper. We also compared the performance of the metric in a GRMHD simulation using the same setup as previous simulations with the matching metric. We found that the superposed metric reproduces the same physical features of the matching metric simulation. We conclude that a superposed metric is a robust approach for exploring the MHD plasma on BBH systems.

Based on these results, in an upcoming paper, we will analyze the effects of the BH spins and orbital evolution on the mini-disk dynamics and outflows of a BBH system embedded in a circumbinary disk \cite{combi2021}.  When the black holes rotate, jets of entirely new characteristics may emerge in a BBH system. Any jet launched should have a helical structure with a diameter equal to the major axis of the responsible black hole's orbit. If each black hole produces a jet, the two jets may collide or interact. Because jets are intrinsically unsteady, and the minidisks' mass accretion rates vary with a phase difference of $\simeq \pi$ \cite{bowen2019}, intersection dynamics are expected to be asymmetric in general even if the black holes have the same mass. This would generate a whole range of unexplored phenomena, such as periodic non-thermal flares produced by the collision region \cite{gutierrez2021}.

One of the main advantages of our approach is that it assumes very little of the BBH properties. Although we have restricted to quasi-circular orbits in this work, implementing a general orbital motion is straightforward because the superposition does not assume any symmetries in the trajectories. As long as the trajectories are solutions of the Post-Newtonian equations, the constraints should remain low. On the other hand, even though we assume that the spins of the BHs are aligned to the orbital plane, we can easily generalize this by applying a rotation to the BH term in \eqref{eq-metric} before the boost transformation. Moreover, this rotation can be time-dependent to take into account the Post-Newtonian evolution of spins. In this way, we will be able to describe the metric of precessing binaries approaching merger in all generality. The last two points, however, must be tested in the same way we did here. We leave them for future work. Finally, using the multi-grid Patchwork MHD code \cite{avara2021}, we would have the possibility to handle the entire parameter space of the system for inspiralling binaries in a computationally efficient way.

\begin{figure*}[htb]
  \begin{center}
  \includegraphics[width=.4\linewidth]{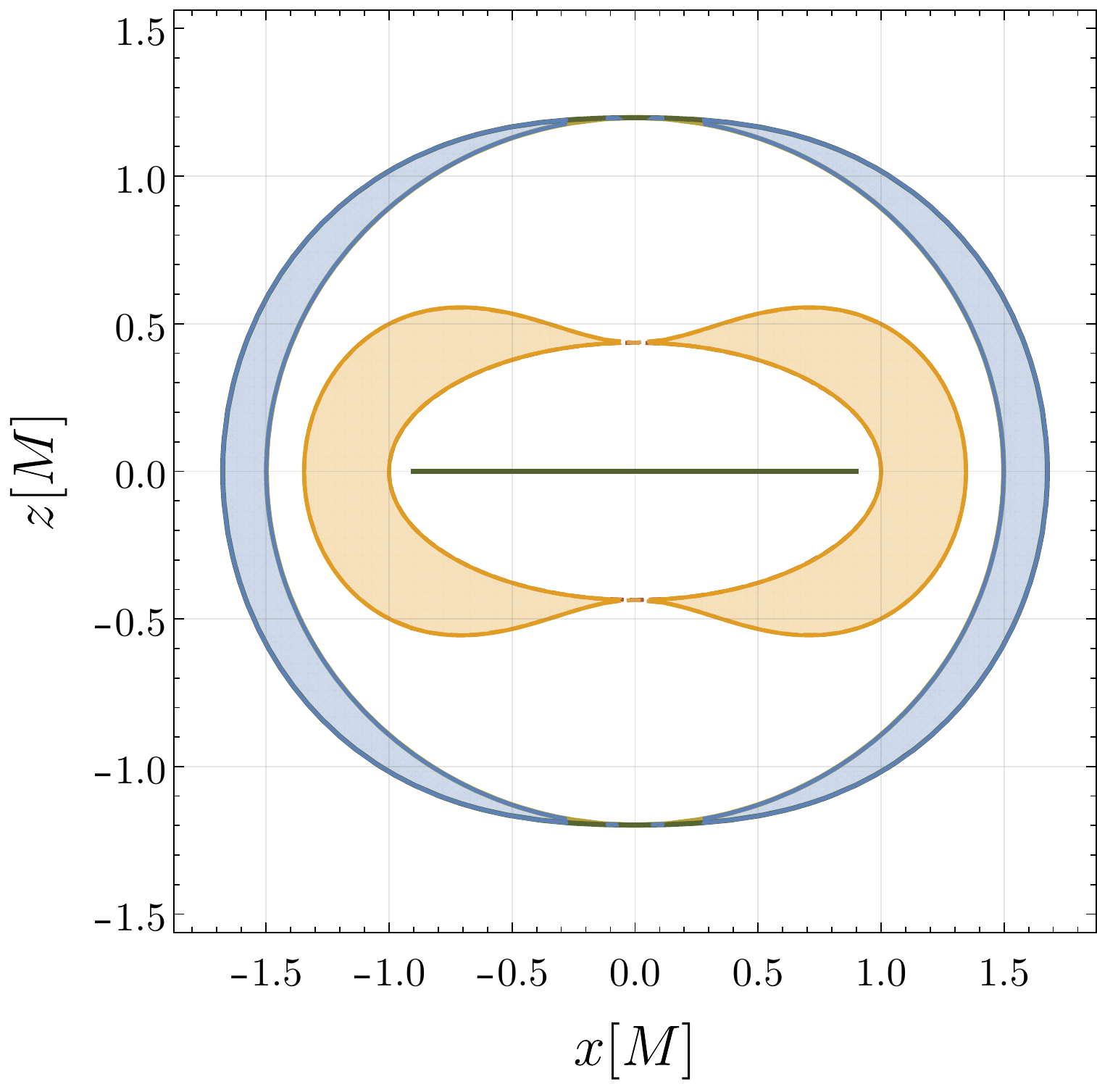}
  \includegraphics[width=.4\linewidth]{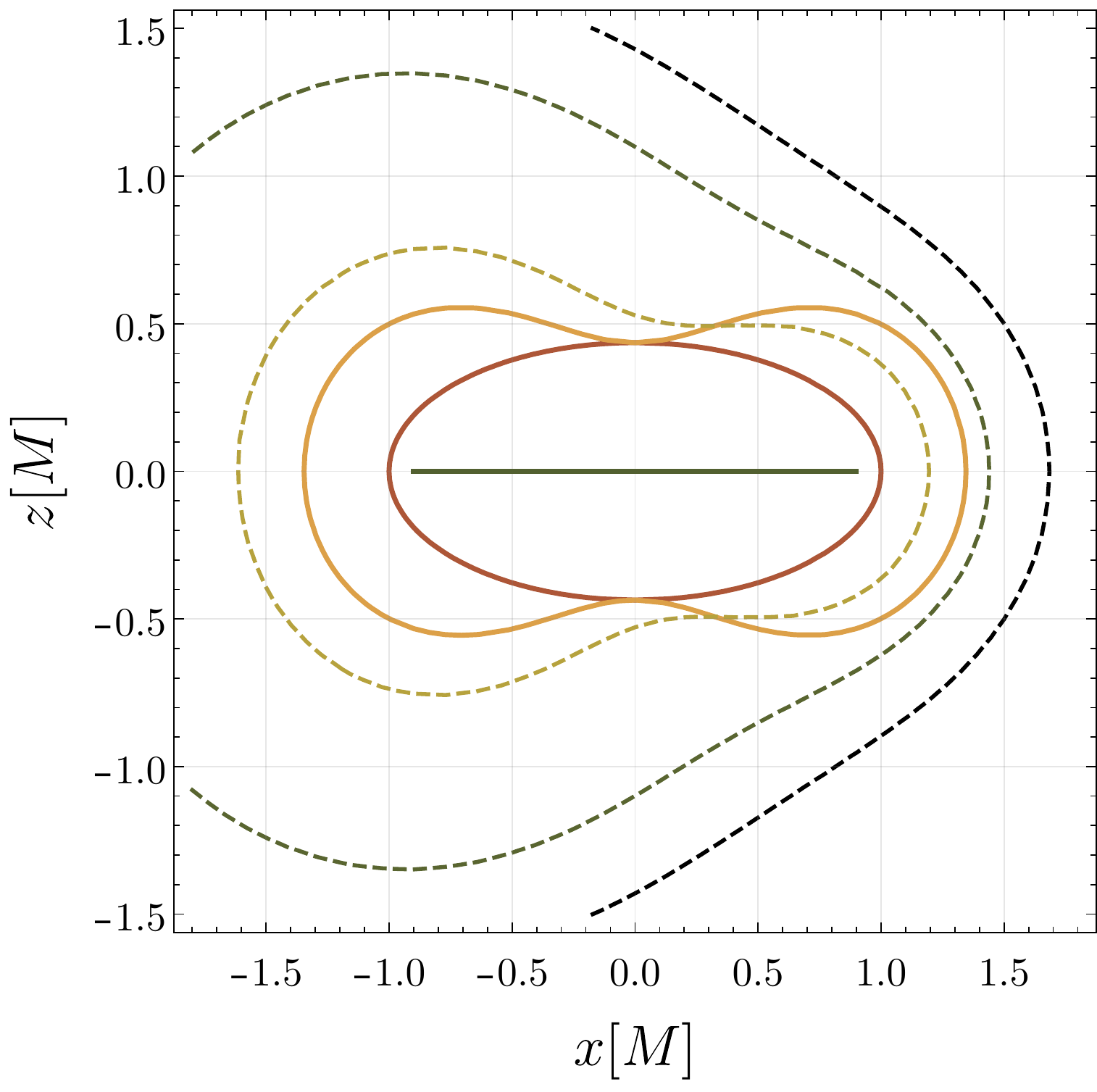}
  \end{center}
  \caption{\textit{Left}: Ergosphere region of Kerr BH with spin $\chi=0.9$ for Kerr-Schild coordinates (blue) and harmonic coordinates (orange). Note that the surfaces in harmonic coordinates are more oblique compared with the Kerr-Schild coordinates. The radius of the singularity (green) is the same for both coordinate systems. \textit{Right}: Ergosphere regions for a x-boosted harmonic Kerr BH with spin $\chi=0.9$ for different velocities ($v/c=0,\,0.1,\,0.5,\,0.9$). The horizon (red) and singularity (green) are the same in each case, but the ergosphere region increases with increasing velocity.}
  \label{fig: horizonergo}
\end{figure*}

\begin{acknowledgments}

We thank Carlos Lousto, Eduardo Guti\'errez, and Mark Avara for useful discussions.
L.~C., F.~L.~A, M.~C., and B.~I acknowledge support from 
 AST-1028087, AST-1028087, AST-1516150, PHY-1707946 
 and from NASA TCAN grant No. 80NSSC18K1488. 
L.~C also acknowledges support from a CONICET (Argentina) fellowship.
S.~C.~N. was supported by AST-1028087, AST-1515982 and
OAC-1515969, and by an appointment to the NASA Postdoctoral Program at
the Goddard Space Flight Center administrated by USRA through a
contract with NASA.
D.~B.~B. is supported by the US Department of Energy through the Los Alamos 
National Laboratory. Los Alamos National Laboratory is operated by Triad National 
Security, LLC, for the National Nuclear Security Administration of U.S. 
Department of Energy (Contract No. 89233218CNA000001).
H.~N. acknowledges support from JSPS KAKENHI Grant Nos. JP16K05347 and JP17H06358.

Computational resources were provided by the Blue Waters
sustained-petascale computing NSF projects OAC-1811228 and OAC-1516125. 
Blue Waters is a joint effort of the University of Illinois at Urbana-Champaign and its
National Center for Supercomputing Applications. Additional resources
were provided by the RIT's BlueSky and Green Pairie Clusters   
acquired with NSF grants AST-1028087, PHY-0722703, PHY-1229173 and PHY-1726215.

The views and opinions expressed in this paper are those of the authors 
and not the views of the agencies or US government.

\end{acknowledgments}



\appendix

\section{Horizons in harmonic coordinates}

To perform GRMHD simulations in a BBH spacetime, we need a clear picture of how the BH singularities and horizons behave in the chosen coordinates. In this appendix, we will show how the harmonic coordinates compare with usual Kerr-Schild coordinates defined in Eq.~\eqref{eq: ks}.

We are interested in the ergosphere $r_{\mathcal{E}}$ and the outer horizon $r_{+}$ of our spacetime. In Boyer-Lindquist (BL) coordinates, these are given by \cite{poisson}:
\begin{equation}
r^{\rm BL}_{\mathcal{E}} = M + \sqrt{M^2-a^2 \cos^2(\theta_{\rm BL})},
\end{equation}
\begin{equation}
r^{\rm BL}_{+} = M + \sqrt{M^2-a^2}.
\end{equation}

The radius in harmonic coordinates can be related with the BL radius, $r_{\rm BL}$, as:
\begin{equation}
r^2_{\rm H} = (r_{\rm BL}-M)^2 + a^2 (1-\cos^2(\theta_{\rm BL}) ),
\end{equation}
where
\begin{equation}
\cos(\theta_{\rm BL}) \equiv \frac{z_{\rm H}}{r_{\rm BL}-M},
\label{eq: blcos}
\end{equation}
and $r_{\rm BL}=r(x_{\rm H},\,y_{\rm H},\,z_{\rm H})$ is calculated from equation \eqref{eq: blradius}. From these expressions, we can derive parametric equations for the horizons and ergosphere in harmonic coordinates:
\begin{equation}
r^{\rm H}_{\mathcal{E}} = M \sqrt{1-\chi^2 \Big(2 \cos^2(\theta_{\rm BL})-1 \Big)},
\end{equation}
\begin{equation}
r^{\rm H}_{+} = M \sqrt{1-\chi^2\cos^2(\theta_{\rm BL})}.
\end{equation}

From Eq.~\eqref{eq: blcos}, we observe that the harmonic coordinates become singular at $r^{\rm BL}_{\mathcal{S}}=M$, which means that there is a disk singularity at $z^{\rm H}=0$ with radius given by:
\begin{equation}
r^{\rm H}_{\mathcal{S}} = M \chi.
\end{equation}

The radius of the horizon in harmonic coordinates shrinks at the poles as $\chi$ increases, while the radius of the horizon at $z=0$ is fixed at $r^{\rm H}_{+}(z=0)=M$ for any value of the spin. Then, the distance between the singularity and the horizon at $z=0$ shrinks with increasing spin as $\delta_{\rm H}:=M(1-\chi)$. In contrast, in Kerr-Schild coordinates, the horizon is further away from the singularity, with a separation of $\delta_{\rm KS}:= \delta_{\rm H} + M g(\chi)$, where $g(\chi):=\sqrt{2+2\sqrt{1-\chi^2}}-1 > 0$ (see Figure \ref{fig: horizonergo}).

To avoid any spurious effect of the inner boundary of the domain, it is usually placed inside the horizon to mask the singularity of the spacetime. In this way, the coordinates must be horizon penetrating and the singularity should be sufficiently far from the horizon. In this regard, the Kerr-Schild coordinate system is more convenient than harmonic coordinates since the distance $\delta_{\rm KS}$ is bigger than $\delta_{\rm H}$. For performing high-spin simulations with the harmonic coordinates, one could artificially remove the singularity by implementing a modification of the metric inside the horizon, e.g., modifying the function $r_{\rm BL}(x_{\rm H})$.

\begin{figure}[h]
  \centering
  \includegraphics[width=.9\linewidth]{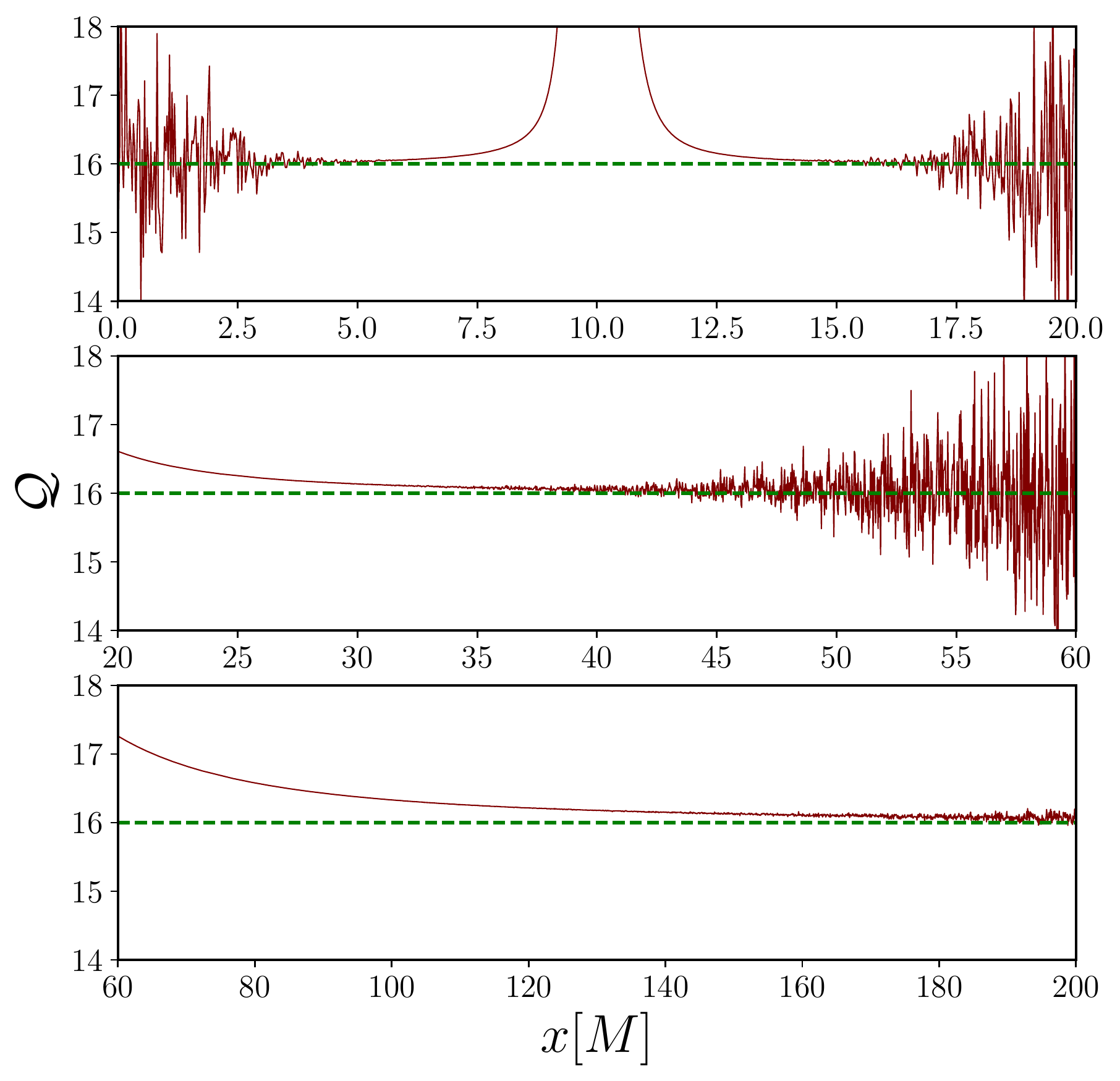}
  \caption{Convergence factor of our numerical scheme for different regions. In the top panel we use $h/M=0.0125$, in the middle panel $h/M=0.1$, and in the bottom panel $h/M=0.8$.}
  \label{fig: convergence}
\end{figure}

Note that the Cartesian Kerr-Schild coordinates used here are not the usual coordinates that accretion disk theorists call `Kerr-Schild' \cite{mckinney2004measurement, kelly2021electromagnetic}. The `accretion-disk Kerr-Schild' coordinates are a modification of the BL coordinates that renders the metric horizon-penetrating but maintains the singularity at $r_{\rm AKS}=r_{\rm BL}=0$. The `Cartesian Kerr-Schild' coordinates that we use here are more common in numerical relativity and appear in the original work of Kerr \cite{visser2007kerr, kerr1963gravitational}.

Finally, let us note that our spacetime contains moving BHs, boosted with respect to the asymptotically flat region. This means that the morphology of the ergosphere would be different from a static BH and these differences can be significant for high velocities. As discussed in Ref. \cite{penna2015}, even a non-spinning BH acquires an ergosphere when the BH is boosted. In the case of a spinning BH, we can see from Figure~\ref{fig: horizonergo} that the ergosphere increases when the BH has higher velocities.

\section{Convergence tests}

Since we are using a finite difference scheme for computing the metric and connection derivatives, we show here the convergence to the analytical solution of the fourth-order discretization. As explained in Ref. \cite{ireland16}, since the metric spans several length scales, we need different mesh spacing to resolve the solution. Given a numerical quantity $U$, we explore the convergence factor, $\mathcal{Q}^h(U)$, defined as:
\begin{equation}
\mathcal{Q}^h(U):= \frac{U^{(4h)}-U^{(2h)}}{U^{(2h)}-U^{(h)}},
\end{equation}
where $h$ is the size of the mesh spacing and $U^{(k)}$ is the numerical approximation with spacing $k$. If the method is well behaved, we have \cite{rezzolla}:
\begin{equation}
\mathcal{Q}^h(U) = 2^p + \mathcal{O}(h).
\label{eq-conv}
\end{equation}

We explore the convergence of the Ricci scalar along the $x$ axis in the equatorial plane, which is the most relevant region. We calculate the derivatives of the metric on a uniform Cartesian grid using a fourth-order finite difference method, so we use $p\equiv4$ in \eqref{eq-conv}. In Figure \ref{fig: convergence}, we show convergence for different regions. In the top panel, for the vicinity of the BH at $[0M,\,20M]$, we use a mesh spacing given by $h/M=0.0125$. Away from the BH, a high-resolution mesh drops the convergence because of the limited machine precision to represent numbers (double precision in our case) \footnote{We can confirm this using a quadrupole precision code, although it is prohibitively expensive for our current simulations \cite{ireland16}}, so we change the mesh to $h/M=0.1$ at $[20M,60M]$ and $h/M=0.8$ at $[60M,200M]$. From Figure \ref{fig: convergence} we can check that convergence is achieved in the different regions, where we expect $\mathcal{Q}\sim 16$.

\bibliographystyle{ieeetr}

\bibliography{bibliography}   

\end{document}